\documentclass{aa}

\usepackage{graphicx,dblfloatfix,fixltx2e}
\usepackage{txfonts}
\usepackage{hyperref}

\begin{document}

   \title{Spatially resolved vertical vorticity in solar supergranulation using helioseismology and local correlation tracking}

\titlerunning{Spatially resolved vertical vorticity in solar supergranulation}
\authorrunning{}

     \author{J. Langfellner \inst{1}
            \and
            L. Gizon \inst{2,1}
            \and
            A.~C. Birch \inst{2}}

    \institute{Georg-August-Universit\"at, Institut f\"ur Astrophysik,
               Friedrich-Hund-Platz 1, 37077 G\"ottingen, Germany
          \and
             Max-Planck-Institut f\"ur Sonnensystemforschung,
             Justus-von-Liebig-Weg 3, 37077 G\"ottingen, Germany
             }

   \date{Received <date> / Accepted <date>}

  \abstract{
   Flow vorticity is a fundamental property of turbulent convection in rotating systems. Solar supergranules exhibit a preferred sense of rotation, which depends on the hemisphere. This is due to the Coriolis force acting on the diverging horizontal flows.
   We aim to spatially resolve the vertical flow vorticity of the average supergranule at different latitudes, both for outflow and inflow regions.
   To measure the vertical vorticity, we use two independent techniques: time-distance helioseismology (TD) and local correlation tracking of granules in intensity images (LCT) using data from the Helioseismic and Magnetic Imager (HMI) onboard the Solar Dynamics Observatory (SDO). Both maps are corrected for center-to-limb systematic errors.
   We find that 8-h TD and LCT maps of vertical vorticity are highly correlated at large spatial scales. Associated with the average supergranule outflow, we find tangential (vortical) flows that reach about $10$~m~s$^{-1}$ in the clockwise direction at $40\degr$ latitude. In average inflow regions, the tangential flow reaches the same magnitude, but in the anti-clockwise direction. These tangential velocities are much smaller than the radial (diverging) flow component (300~m~s$^{-1}$ for the average outflow and $200$~m~s$^{-1}$ for the average inflow). The results for TD and LCT as measured from HMI are in excellent agreement for latitudes between $-60\degr$ and $60\degr$. From HMI LCT, we measure the vorticity peak of the average supergranule to have a full width at half maximum of about 13~Mm for outflows and 8~Mm for inflows. This is larger than the spatial resolution of the LCT measurements (about 3~Mm). On the other hand, the vorticity peak in outflows is about half the value measured at inflows (e.g.~$4\times10^{-6}$~s$^{-1}$ clockwise compared to $8\times10^{-6}$~s$^{-1}$ anti-clockwise at $40\degr$ latititude).
Results from the Michelson Doppler Imager (MDI) onboard the Solar and Heliospheric Observatory (SOHO) obtained in 2010 are biased compared to the HMI/SDO results for the same period.
}

   \maketitle
%


\section{Introduction}
\citet{duvall_2000} and \citet{gizon_2003} revealed that supergranules \citep[see][for a review]{rieutord_2010} possess a statistically preferred sense of rotation that depends on solar latitude. In the northern hemisphere, supergranules tend to rotate clockwise, in the southern hemisphere anti-clockwise. This is due to the Coriolis force acting on the divergent horizontal flows of supergranules. For supergranulation (lifetime $>1$~day), the Coriolis number is close to unity \citep[see][]{gizon_2010}. As a consequence, the vorticity induced by the Coriolis force should be measurable by averaging the vorticity of many realizations of supergranules at a particular latitude.

For single realizations, \citet{attie_2009} detected strong vortices associated with supergranular inflow regions by applying a technique called balltracking. \citet{komm_2007} presented maps of vortical flows in quiet Sun convection using helioseismic ring-diagram analysis. With the same technique, \citet{hindman_2009} resolved the circular flow component associated with inflows into active regions.
The spatial structure of such vortical flows has not yet been studied though for many realizations. Knowledge of the flow structure of the average supergranule will help constrain models and simulations of turbulent convection that take into account rotation.

Here, we aim to spatially resolve the vertical component of flow vorticity associated with the average supergranule. We investigate both outflows from supergranule centers and inflows into the supergranular network.
To measure the flow divergence and vorticity, we use two independent techniques: time-distance helioseismology (TD) and local correlation tracking (LCT) of granules. We use the TD method from \citet{langfellner_2014}, where a measurement geometry that is particularly sensitive to the vertical component of flow vorticity was defined.

\subsection{Time-distance helioseismology}
Time-distance helioseismology makes use of waves travelling through the Sun \citep{duvall_1993}. A wave travelling from the surface point $\vec{r}_1$ through the solar interior to another surface point $\vec{r}_2$ is sensitive to local physical conditions (e.\,g.~the wave speed or density). A flow in the direction $\vec{r}_2 - \vec{r}_1$ will increase the wave speed, thus reducing the travel time $\tau^+$ from $\vec{r}_1$ to $\vec{r}_2$. A flow in the opposite direction will result in a longer travel time. The travel time is measured from the temporal cross-covariance, labeled $C$, of the observable $\phi$ obtained at the points $\vec{r}_1$ and $\vec{r}_2$:
\begin{equation}
 C(\vec{r}_1,\vec{r}_2,t) = \frac{h_t}{T}\sum_{i=-N}^N \phi(\vec{r}_1,t_i)\phi(\vec{r}_2,t_i+t),
\end{equation}
where $h_t$ is the temporal cadence, $T=2(N+1)h_t$ is the observation time, and $t_i = (-N,-N+1,\dots,N) h_t$ are the times when $\phi$ is sampled. Typically, the observable $\phi$ is the Doppler line-of-sight velocity component.

The travel time can be obtained from $C$ by fitting a wavelet \citep{duvall_1997} or by comparison with a reference cross-covariance and application of an appropriate weight function \citep{gizon_2004}. To distinguish the flow signal in the travel time from other perturbations (e.\,g. local sound speed changes), we use the travel-time difference
\begin{equation}
\tau^\text{diff}(\vec{r}_1,\vec{r}_2) = \tau^+(\vec{r}_1,\vec{r}_2) - \tau^+(\vec{r}_2,\vec{r}_1) .
\end{equation}

\begin{figure}
  \centering
 \includegraphics[width=0.4\hsize]{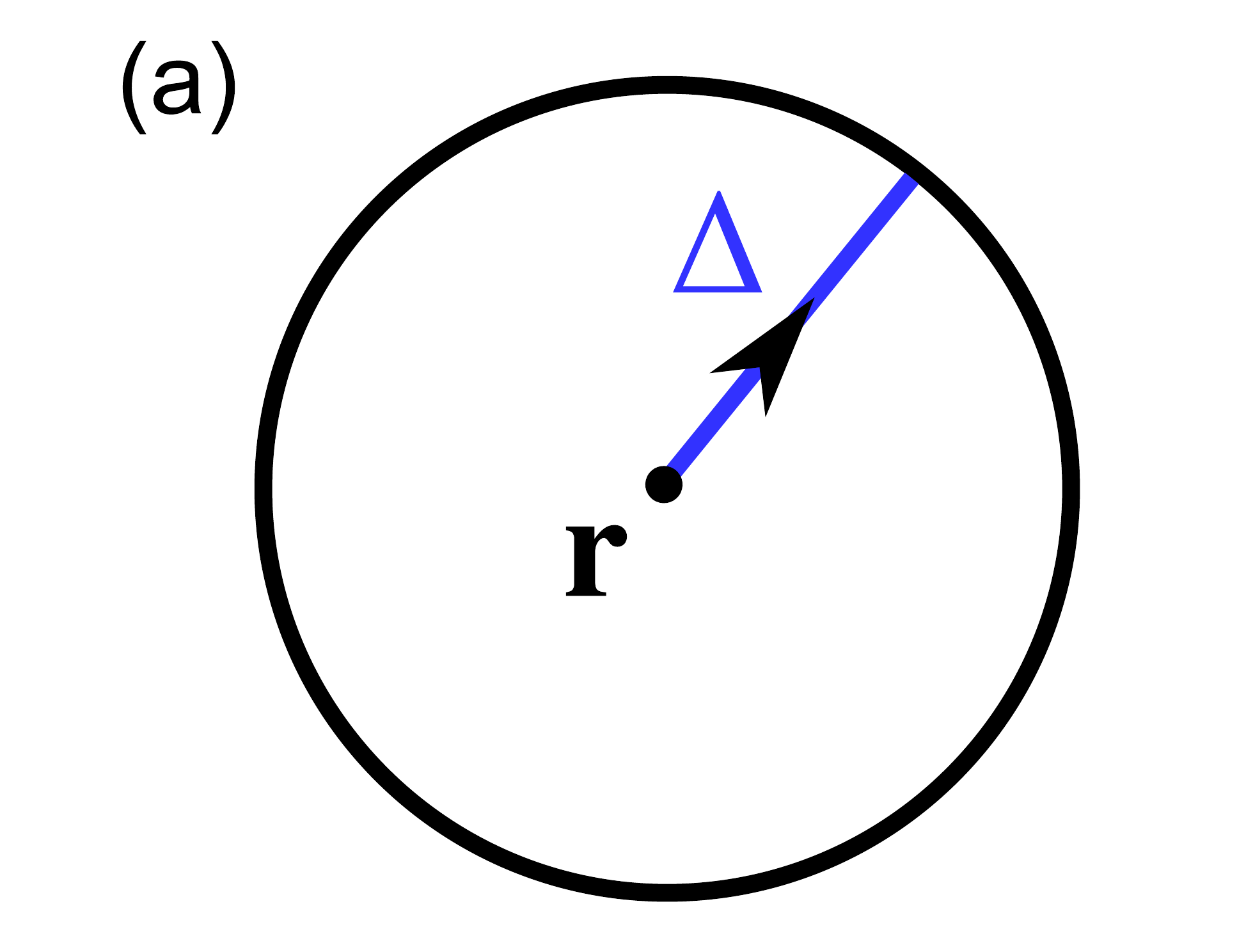}
 \includegraphics[width=0.45\hsize]{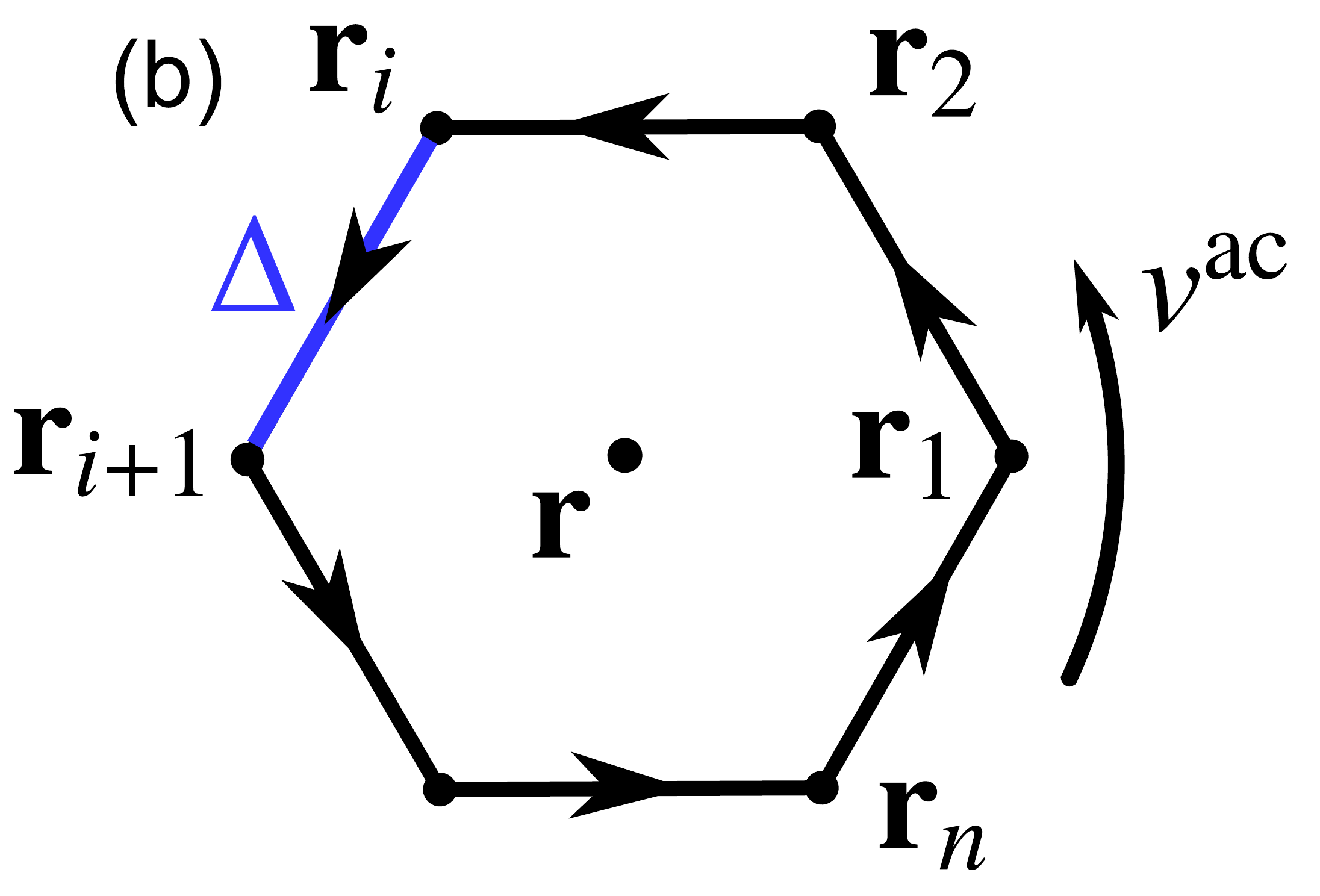} \\ \vspace{0.5cm}
 \includegraphics[width=0.4\hsize]{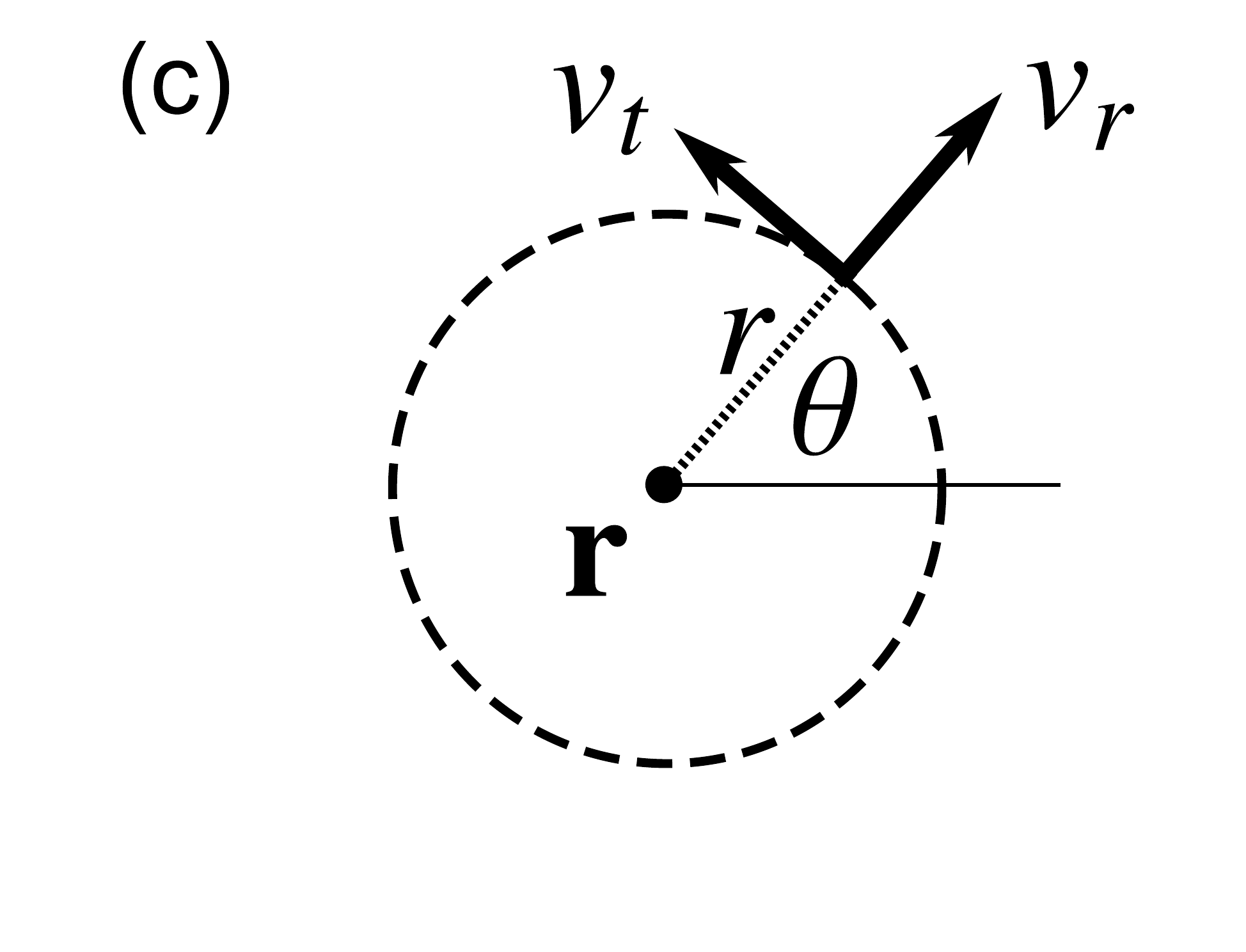}
 \includegraphics[width=0.45\hsize]{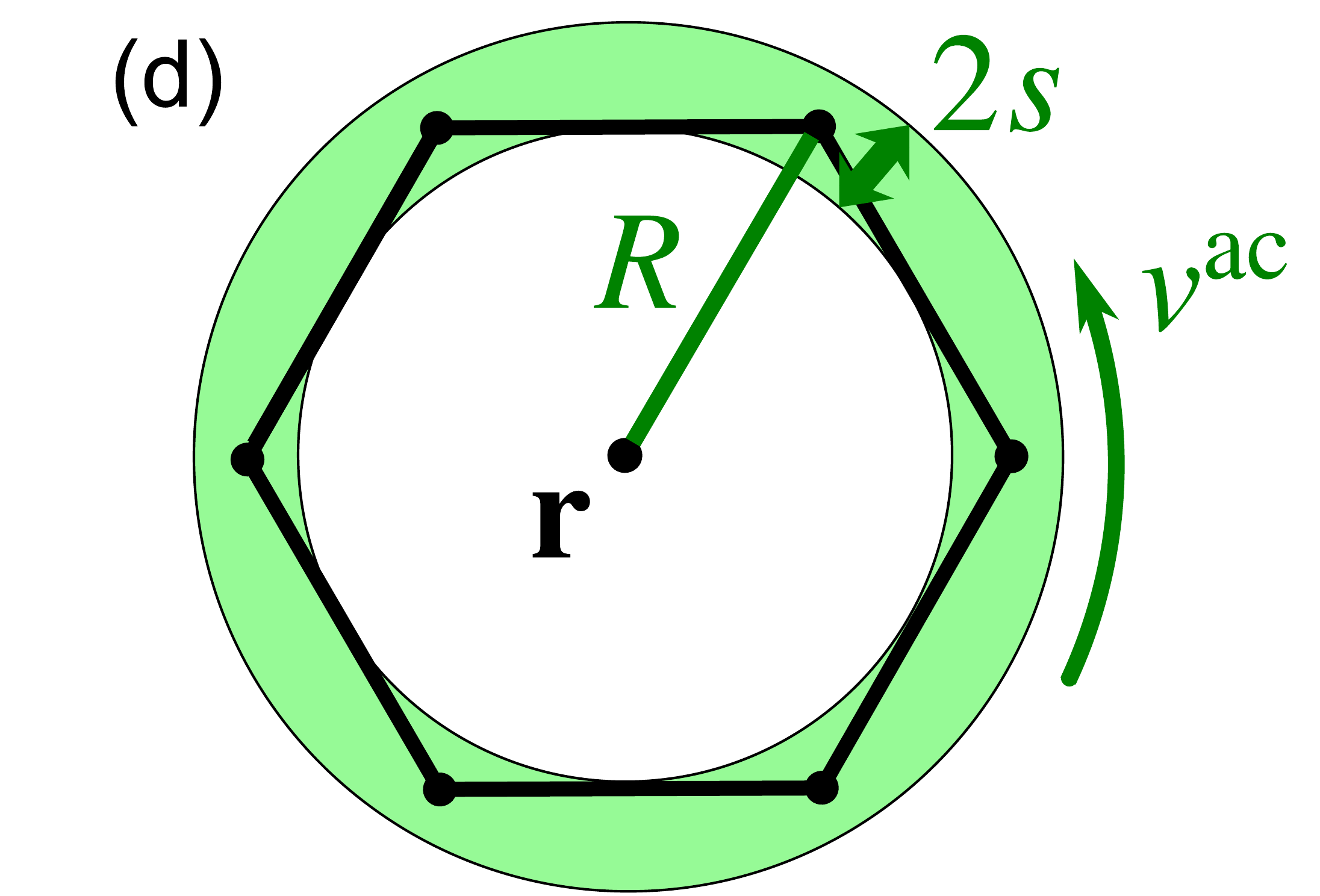}
   \caption{Measurement geometries for divergence- and vorticity-sensitive travel times and LCT velocities. \textbf{a)} The divergence-sensitive travel time $\tau^\text{oi}$ is obtained by measuring the travel-time difference between a central point $\vec{r}$ and a surrounding annulus of radius $\Delta$. \textbf{b)} The vorticity-sensitive travel time $\tau^\text{ac}$ is obtained by measuring the travel-time differences $\tau^\text{diff}$ between adjacent points along a regular polygon surrounding $\vec{r}$. The polygon consists of $n$ points and has edges of length $\Delta$. The points are situated on a circle of radius $R = \Delta/[2 \sin(\pi/n)]$. The travel time $\tau^\text{ac}$ is the average over the components $\tau^\text{diff}$ (see Eq.~(\ref{eq_tauccw-def})). We obtain circulation velocities $v^\text{ac}$ from multiplying $\tau^\text{ac}$ by a calibration factor. \textbf{c)} From LCT, we obtain the horizontal velocity components $v_x$ and $v_y$. Given a reference point $\vec{r}$, the LCT velocities can be expressed in 2D polar coordinates $(r,\theta)$ by the outward pointing radial velocity component $v_r$ and the anti-clockwise pointing tangential velocity component $v_t$. \textbf{d)} Using LCT velocities, we approximate $v^\text{ac}$ by averaging the tangential velocity component $v_t$ over the annulus shaded in green. The annulus is defined by its radius $R$ and half-width $s$. We choose $R = \Delta$ for $n = 6$ and $s = 2~$Mm.}
\label{fig_geometry}
    \end{figure}

Travel times that are especially sensitive to the horizontal flow divergence can be obtained by replacing $\vec{r}_2$ with an annulus around $\vec{r}_1$ (see Fig.~\ref{fig_geometry}a). Averaging $\phi$ over the annulus yields the ``outward--inward'' travel time $\tau^\text{oi}$ \citep{duvall_1996}. To obtain travel times that measure the vertical component of the flow vorticity, we average $\tau^\text{diff}$ components along a closed contour in anti-clockwise direction \citep{langfellner_2014}. We choose the contour to be a regular polygon with $n$ points and edge length $\Delta$ in order to approximate an annulus (see Fig.~\ref{fig_geometry}b). The mean over the $\tau^\text{diff}$ components gives the vorticity-sensitive $\tau^\text{ac}$ travel time:
\begin{equation}
 \tau^\text{ac}(\vec{r},\Delta,n) := \frac{1}{n} \sum_{i=1}^n \tau^\text{diff}(\vec{r}_i,\vec{r}_{i+1}) \ ,
 \label{eq_tauccw-def}
\end{equation}
where we use the notation $\vec{r}_{n+1} = \vec{r}_1$.

\subsection{Local correlation tracking}
LCT measures how structures in solar images are advected due to background flows. For the tracer, it is common to use solar granulation observed in photospheric intensity images \citep{november_1988}. The general procedure is as follows. Pairs of images are selected that observe the same granules but are separated by a time $\Delta t$. This time separation must be small compared to the lifetime of granules, i.e. $\Delta t \ll 10~$min. To obtain spatially resolved velocity maps, an output map grid is defined. For each grid point, subsets of the intensity images that are centered around the grid point are selected by applying a spatial window and multiplied by a Gaussian with a full width at half maximum (FWHM) of typically a few megameters. The subsets are then cross-correlated in the two spatial image dimensions $x$ and $y$. The peak position $(\Delta x,\Delta y)$ of the cross-correlation yields the spatial shift. Since the measured shift is usually only a small fraction of a pixel, it must be obtained using an appropriate fitting procedure. Finally, the velocity components in $x$ and $y$ directions are given by $v_x = \Delta x/\Delta t$ and $v_y = \Delta y/\Delta t$.

The LCT method has proven valuable to measure flow patterns in the Sun. For instance, \citet{brandt_1988} and \citet{simon_1989} observed single vortex flows at granulation scale. \citet{hathaway_2013} detected giant convection cells with LCT of supergranules in Doppler velocity images. For a comparison of different LCT techniques, see \citet{welsch_2007}.


\section{Observations and data processing}  \label{sect_observations}
The basis for our measurements of wave travel times and flow velocities from local correlation tracking are two independent observables. We use Doppler velocity images for the TD and intensity images for the LCT. Both observables are measured for the full solar disk by the Helioseismic and Magnetic Imager (HMI) onboard the Solar Dynamics Observatory (SDO) \citep{schou_2012} and are available for the same periods of time. This allows a direct comparison of the two methods for looking at ``the same Sun'' but utilizing independent data.

We used 112 days of both SDO/HMI Dopplergrams and intensity images in the period from 1 May through 28 August 2010. Patches of approximate size $180\times180$~Mm$^2$ were selected that are centered at solar latitudes from $-60\degr$ to $60\degr$ in steps of $20\degr$. They were tracked for 24~h each at a rate consistent with the solar rotation rate from \citet{snodgrass_1984} at the center of the patch. The data cubes cross the central meridian approximately at half the tracking time. They were remapped using Postel's projection with a spatial sampling of 0.5~arcsec px$^{-1}$ (0.348~Mm px$^{-1}$). The temporal cadence is 45~s. We divided each data cube into three 8~h datasets. The $x$ direction of the remapped images points to the west, the $y$ direction points to the north.

For further comparison, we also used Dopplergrams from the Michelson Doppler Imager (MDI) onboard the Solar and Heliospheric Observatory (SOHO) spacecraft \citep{scherrer_1995}. We chose 59 days of images taken in the MDI full-disk mode that overlap in time with the HMI data (8 May through 11 July 2010). We tracked and remapped the MDI Dopplergrams in the same manner as for HMI, albeit with a coarser spatial sampling of 2.0~arcsec px$^{-1}$ (1.4~Mm px$^{-1}$).

\subsection{Flow velocity maps from local correlation tracking}
   \begin{figure*}
  \centering
\includegraphics[width=0.33\hsize]{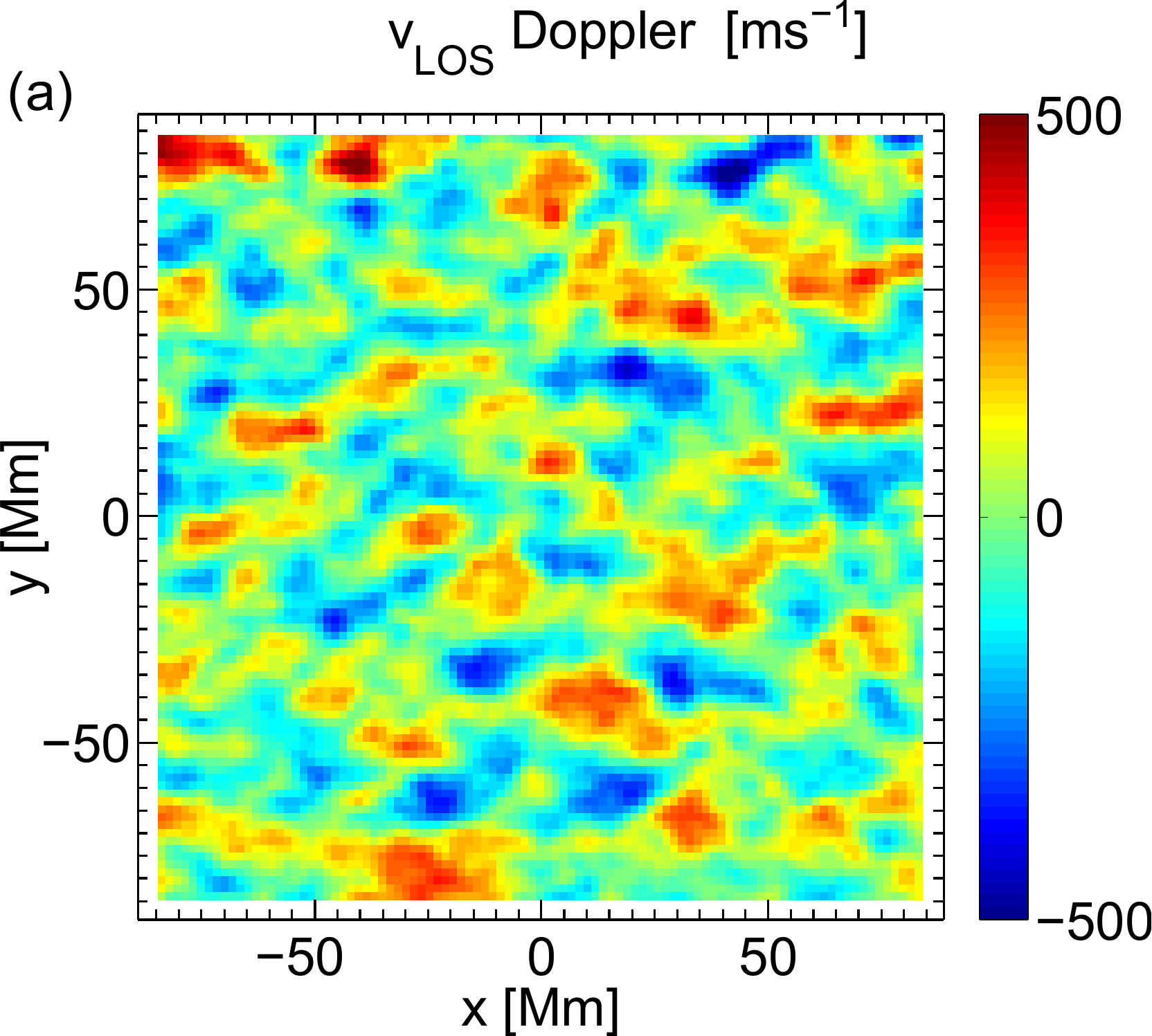}
\includegraphics[width=0.33\hsize]{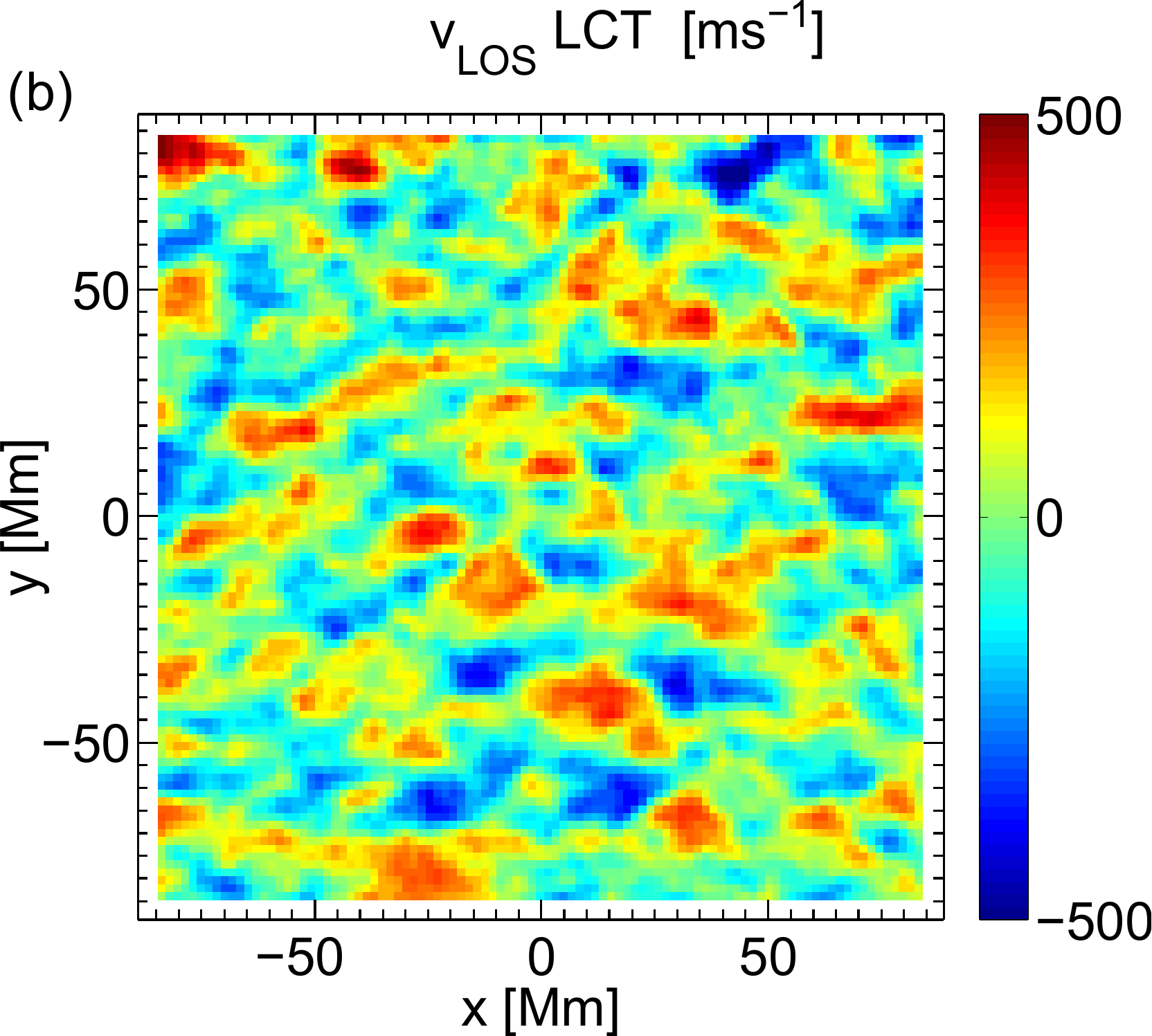}
\includegraphics[width=0.32\hsize]{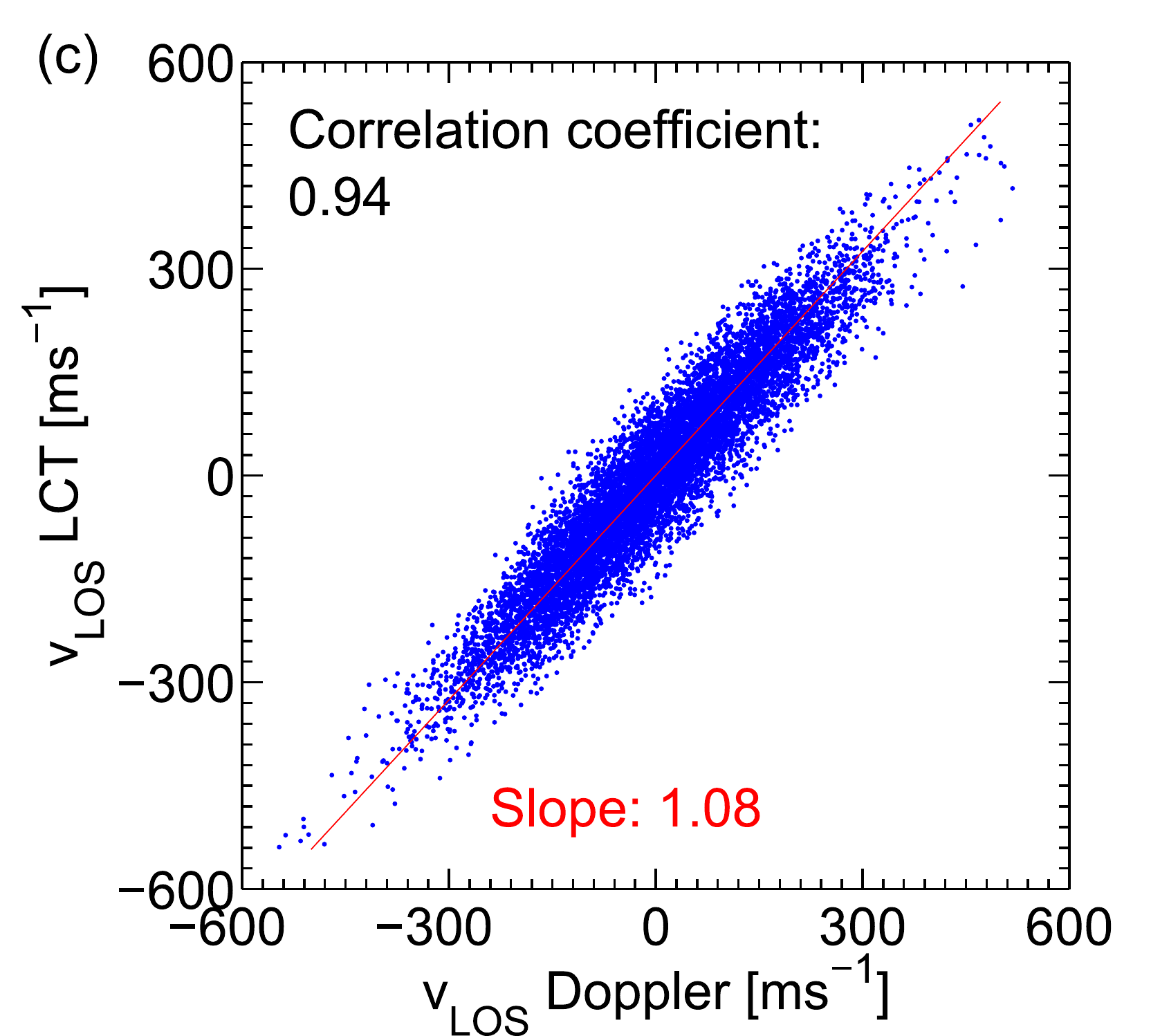}
   \caption{Comparison of line-of-sight velocity from two different data products at $40\degr$ solar latitude on 6 June 2010. \textbf{a)} HMI Dopplergram averaged over 8~h. The map has been convolved with a Gaussian of $\sigma/\sqrt{2} \approx 1.4$~Mm and subsampled to match the coarser LCT resolution. The mean over the map and a linear function in the $x$ direction (parameters determined by a least-squares fit) have been subtracted. \textbf{b)} LCT map from HMI intensity images, averaged over 8~h. The line-of-sight velocity component was computed from the $v_x$ and $v_y$ components. For $v_y$, the mean over the map and a linear function in $y$ direction have been subtracted. \textbf{c)} Scatter plot of the two maps. The Pearson correlation coefficient is 0.94. The red line shows the direction of largest scatter and crosses the origin. It is a best-fit line in the sense that it minimizes the sum of squared distances of the points perpendicular to the line \citep[cf.][]{pearson_1901}. This is different from linear regression, where no error in the $x$ coordinate is assumed and only the sum of squared distances in the $y$ coordinate is minimized. The slope of the red line is $1.08$; the error in the direction of lowest scatter is $33.9$~m~s$^{-1}$.}
\label{fig_directdoppler_lct}
    \end{figure*}

For the LCT, we used our own code with the HMI photospheric intensity images as input. Our code is similar to the FLCT code by \citet{fisher_2008}, but uses another procedure to measure the peak positions of the cross-correlation (described later in this section).

We removed the temporal mean image for every dataset and chose an output grid with a sampling of 2.5~arcsec~px$^{-1}$ (1.7~Mm~px$^{-1}$), thus five times coarser than the input images. The size of the image subsets used for the cross-correlation is adapted to the width of the Gaussian the subsets are multiplied by. We chose $\sigma = 2$~Mm for the Gaussian and a diameter of $4\sigma = 8$~Mm for the subsets both in $x$ and $y$ directions. The subsets are separated in time by $\Delta t = 45~$s (the cadence), which is sufficiently small compared to the granules' evolution timescale.
We averaged the cross-correlations over the whole 8~h dataset.

To measure the peak position $(\Delta x, \Delta y)$ of the cross-correlation, we calculated (separately for $x$ and $y$ directions) the parameters of a parabola matching the cross-correlation at the maximum and the adjacent pixels. To improve the estimate of the peak position, we translated the cross-correlation by $(-\Delta x, -\Delta y)$ using Fourier interpolation and iterated the parabolic fit. We repeated this procedure four times in total. The measured shifts converge quickly, the maximum additional shift in a fifth iteration is of the order $10^{-5}$~px at $60\degr$ latitude (corresponding to $0.2~$m~s$^{-1}$ or less), the root mean square of the additional velocity shift is less than $0.02~$m~s$^{-1}$. The measured peak position is the sum of the shifts measured in each step.

In Fig.~\ref{fig_directdoppler_lct}, we compare the line-of-sight component $v_\text{LOS}$ of the LCT velocity with the velocity of an average Dopplergram, obtained by averaging Dopplergrams over the same time-period as the LCT maps (8~hours). Both images are for the same region at $40\degr$ latitude around the central meridian. At this latitude, the average Dopplergrams are dominated by the horizontal flows that can be measured with LCT, but systematic effects like foreshortening are weak (see Appendix~\ref{sect_center-to-limb-systematics}). We convolved the average Dopplergram with a Gaussian of width $\sigma/\sqrt{2} \approx 1.4$~Mm (FWHM roughly 3~Mm). This resembles the convolution of the intensity maps prior to computing the correlation of image subsets in the LCT. The chosen width maximizes the correlation coefficient between the average Dopplergram and the LCT image. In addition, we interpolated the average Dopplergram onto the coarser LCT grid and at each pixel subtracted the mean velocity over the map. To remove the residual rotation signal, we further subtracted a linear gradient in the $x$ direction that we obtained from a least-squares fit of $v_x$ averaged over $y$. The LCT line-of-sight velocity component has been computed from $v_x$ and $v_y$. The $v_y$ map showed a linear gradient in the $y$ direction leading to an average velocity difference of about $-200$~m~s$^{-1}$ between the bottom and the top of the map. This gradient is presumably due to the ``shrinking Sun'' effect, which has been discussed in \citet{lisle_2004}, albeit for LCT of Dopplergrams (a short description is also given in Appendix~\ref{sect_center-to-limb-systematics}). The gradient (and the mean over the $v_y$ map) was removed before computing $v_\text{LOS}$.

The processed $v_\text{LOS}$ maps from direct Doppler data and LCT agree well (correlation coefficient 0.94). The scatter plot shows that on average the velocity values from LCT are slightly larger (by a factor 1.08) than from the Dopplergrams. This is different from what other authors have reported. \citet{derosa_1998} measured a slope of 0.89 and later \citep{derosa_2004} 0.69 using their LCT code and SOHO/MDI Dopplergrams. \citet{rieutord_2001} and \citet{verma_2013} found that LCT underestimates the real velocities in convection simulations (however at smaller spatial scales than we are studying).

\subsection{Travel-time maps for horizontal divergence and vertical vorticity}
As input for the travel-time measurements, we used the HMI Dopplergrams. In Fourier space, we filtered the 8~h datasets to select either the f-mode or p$_1$-mode ridge. The filters consist of a raised cosine function with a plateau region in frequency around the ridge maximum for every wavenumber $k$. Additionally, power for $kR_\odot < 300$ and $kR_\odot > 2\,600$ (f modes) and for $kR_\odot < 180$ and $kR_\odot > 1\,800$ (p$_1$ modes) respectively is discarded. The symbol $R_\odot$ denotes the solar radius. The filter details are given in Appendix \ref{sect_filters}.

We computed the cross-correlation $C$ from each filtered Doppler dataset in temporal Fourier space using
\begin{equation}
 C(\vec{r}_1,\vec{r}_2,\omega) = h_\omega \phi_T^*(\vec{r}_1,\omega) \phi_T(\vec{r}_2,\omega),
\end{equation}
where $\omega$ denotes the angular frequency, $h_\omega = 2\pi/T$ (with $T = 8$~h) is the frequency resolution, and $\phi_T$ is the temporal Fourier transform of the filtered dataset (multiplied by the temporal window function).

For each dataset, we measured travel times $\tau^\text{oi}$ with an annulus radius of $10$~Mm and $\tau^\text{ac}$ with the parameters $\Delta = 10$\,Mm and $n=6$ (regular hexagon). We rotated the hexagon structure successively three times by an angle of $15\degr$ to obtain four $\tau^\text{ac}$ measurements for the same dataset that are only weakly correlated \citep[see][for details]{langfellner_2014}. Averaging over these measurements yields a higher signal-to-noise ratio (S/N). We used the linearized travel-time definition by \citet{gizon_2004} and a sliding reference cross-covariance that we obtained by averaging $C$ over the entire map.

   \begin{figure*}
  \sidecaption
\includegraphics[width=0.34\hsize]{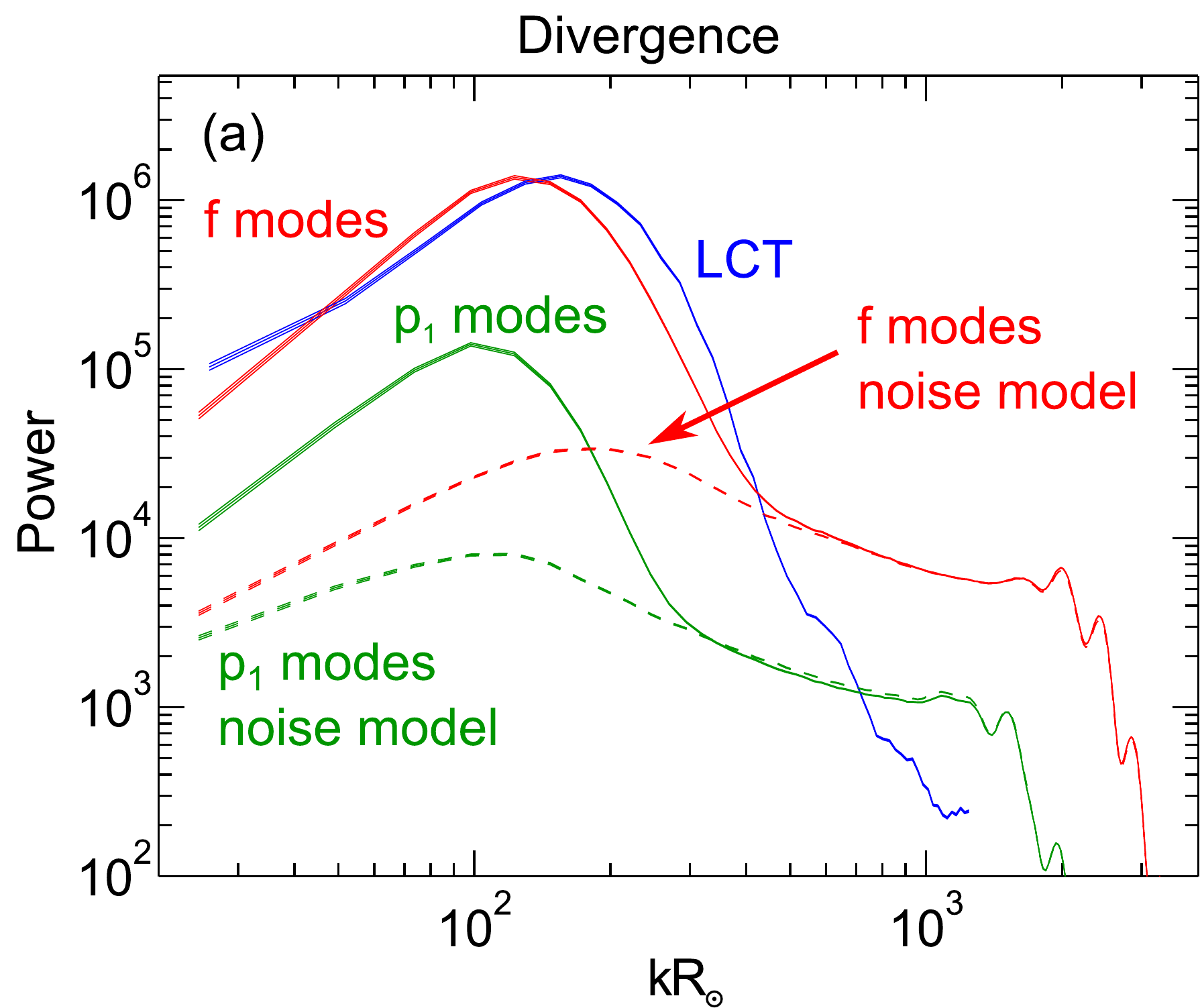}
\includegraphics[width=0.34\hsize]{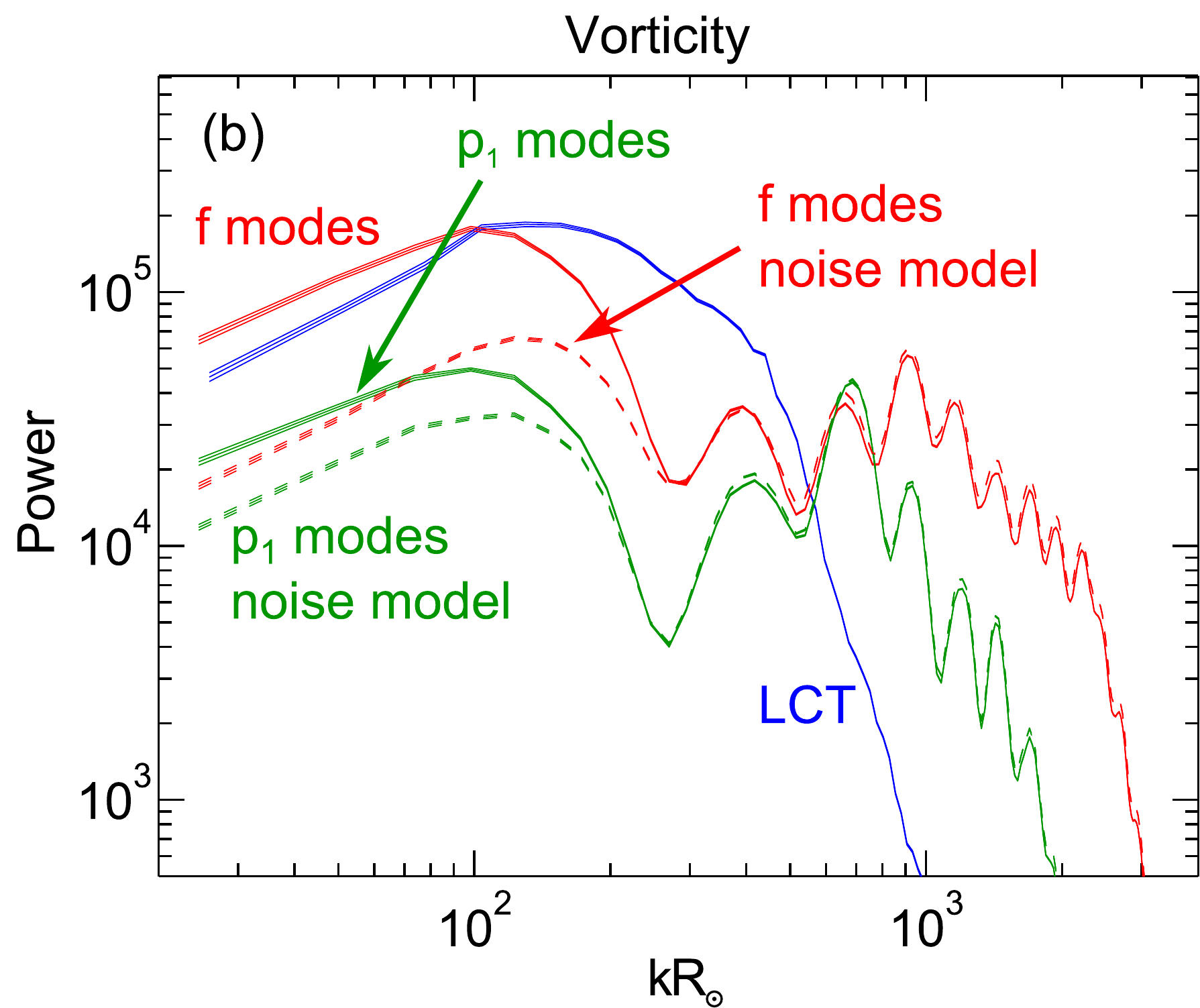}
   \caption{Power spectra (averaged over azimuth of wavevector $\vec{k}$ and 336 datasets) of TD travel-time maps and LCT $\text{div}_h$ and $\omega_z$ maps computed from HMI Dopplergrams and intensity images at the solar equator near disk center (distributed symmetrically between $7\degr$ east and west of the central meridian). \textbf{a)} Divergence-sensitive travel times $\tau^\text{oi}$ for f modes and p$_1$ modes as well as LCT $\text{div}_h$. \textbf{b)} Vorticity-sensitive travel times $\tau^\text{ac}$ for f modes and p$_1$ modes as well as LCT $\omega_z$. The amplitudes of LCT $\text{div}_h$ and $\omega_z$ power were rescaled to match the range of the travel-time power. The dashed lines represent noise models for the f and p$_1$ modes based on \citet{gizon_2004}. The thickness of the lines denotes the $1\sigma$ error.}
\label{fig_power}
    \end{figure*}

\section{Comparison of horizontal divergence and vertical vorticity from TD and LCT}
We now want to compare the measurements of horizontal divergence and vertical vorticity from TD and LCT for HMI. For TD, we use $\tau^\text{oi}$ and $\tau^\text{ac}$ as divergence- and vorticity-sensitive quantities. For LCT, we can directly compute $\text{div}_h = \partial_x v_x + \partial_y v_y$ and $\omega_z = \partial_x v_y - \partial_y v_x$ from the $v_x$ and $v_y$ maps. To compute the derivatives of the LCT velocities, we apply Savitzky-Golay filters \citep{savitzky_1964} for a polynomial of degree three and a window length of 15 pixels (about 5~Mm, with a FWHM of about 3~Mm of the smoothing kernel). The Savitzky-Golay filters smooth out variations in the derivatives on spatial scales below the LCT resolution.

In the case of vorticity, we can also attempt a more direct comparison of TD and LCT. Consider the horizontal velocity field in 2D polar coordinates around $\vec{r}$ (see Fig.~\ref{fig_geometry}c). Instead of using $v_x$ and $v_y$ for LCT, we can study the velocity component in the radial (divergent) direction, $v_r$, and the component in the tangential (anti-clockwise) direction, $v_t$. The travel time $\tau^\text{ac}$ essentially measures $v_t$ averaged over the closed contour. The travel-time $\tau^\text{ac}$ is built up of point-to-point components $\tau^\text{diff}$ that capture the flow component that is parallel to the line connecting the two measurement points. The velocity magnitude that corresponds to the travel time $\tau^\text{diff}$ can roughly be estimated by calibration measurements using a uniform flow (Appendix \ref{chap_conversion}). We use this calibration to convert $\tau^\text{ac}$ travel times into flow velocities and call the result $v^\text{ac}$. 
Since convective flows are highly turbulent, a conversion factor obtained from uniform flows has to be treated with caution though. Additionally, the conversion factor is sensitive to the details of the ridge filter (Appendix \ref{sect_filter-impact}). Also note that because no inversion is applied, the velocities $v^\text{ac}$ represent an average over a depth range given by travel-time sensitivity kernels. For f modes, the range is from the surface to a depth of about 2~Mm, with a maximum of sensitivity near the surface, and for p$_1$ modes from the surface to roughly 3~Mm, with one maximum near the surface and another one at a depth of about 2~Mm \citep[see, e.g.,][]{birch_2007}.
With LCT, we approximate $v^\text{ac}$ by averaging $v_t$ over an hard-edge annulus with radius $R = 10~$Mm and half-width $s = 2~$Mm (see Fig.~\ref{fig_geometry}d). The annulus width roughly corresponds to the width of travel-time sensitivity kernels \citep[see, e.g.,][]{jackiewicz_2007}.

For the divergence-sensitive measurements, such a comparison is not possible without an inversion of the $\tau^\text{oi}$ maps. Therefore, we stick to comparing TD $\tau^\text{oi}$ and LCT $\text{div}_h$ in the following.

\subsection{Spatial power spectra of horizontal divergence and vertical vorticity}
From the TD $\tau^\text{oi}$ and $\tau^\text{ac}$ maps as well as the LCT $\text{div}_h$ and $\omega_z$ maps, we calculated the spatial power spectra and averaged them over azimuth. The result for HMI is shown in Fig.~\ref{fig_power}. Note that we rescaled the amplitude of the LCT power in order to show it together with the travel-time power. 

For the divergence, the TD and LCT powers show a similar behavior at larger scales (except for $kR_\odot = 25$, which corresponds to the map size). However, all three curves peak at different scales -- f modes at $kR_\odot = 120$, p$_1$ modes at $kR_\odot = 100$ and LCT at $kR_\odot = 150$. The comparison with the curves for the TD noise model \citep{gizon_2004} shows that the highest S/N for the TD $\tau^\text{oi}$ occurs at supergranulation scale, with p$_1$ modes probing slightly larger scales than f modes. For LCT, no noise model is available that we know of. Thus it remains unclear if the peak of the power coincides with the peak of the S/N. For small scales ($kR_\odot$ larger than 300) the LCT power vanishes quickly, whereas the TD power reaches a noise plateau (f at $kR_\odot = 500$, p$_1$ at $kR_\odot = 300$).

In the case of vorticity, the curves for TD and LCT look similar at large scales, albeit the power for LCT $\omega_z$ drops more quickly toward larger scales than for TD $\tau^\text{ac}$. Compared to the divergence case, the peak positions are slightly shifted toward larger scales. However, the comparison with the TD noise model reveals that the S/N does not have a peak at supergranulation scale but continues to increase toward larger scales \citep[cf.][]{langfellner_2014}. At mid scales, the LCT power drops off only slowly, whereas the TD power quickly reaches the noise level (f at $kR_\odot = 250$, p$_1$ at $kR_\odot = 200$). It is not clear if the considerably larger power of LCT $\omega_z$ at mid scales ($150 < kR_\odot < 500$) is due to real flows or noise. At smaller scales, both TD curves behave more erratically. This happens, however, in a regime of almost pure noise. LCT power drops off quickly beyond $kR_\odot = 400$.

\subsection{Maps of horizontal divergence and vertical vorticity}
   \begin{figure*}
  \sidecaption
\includegraphics[width=12.5cm]{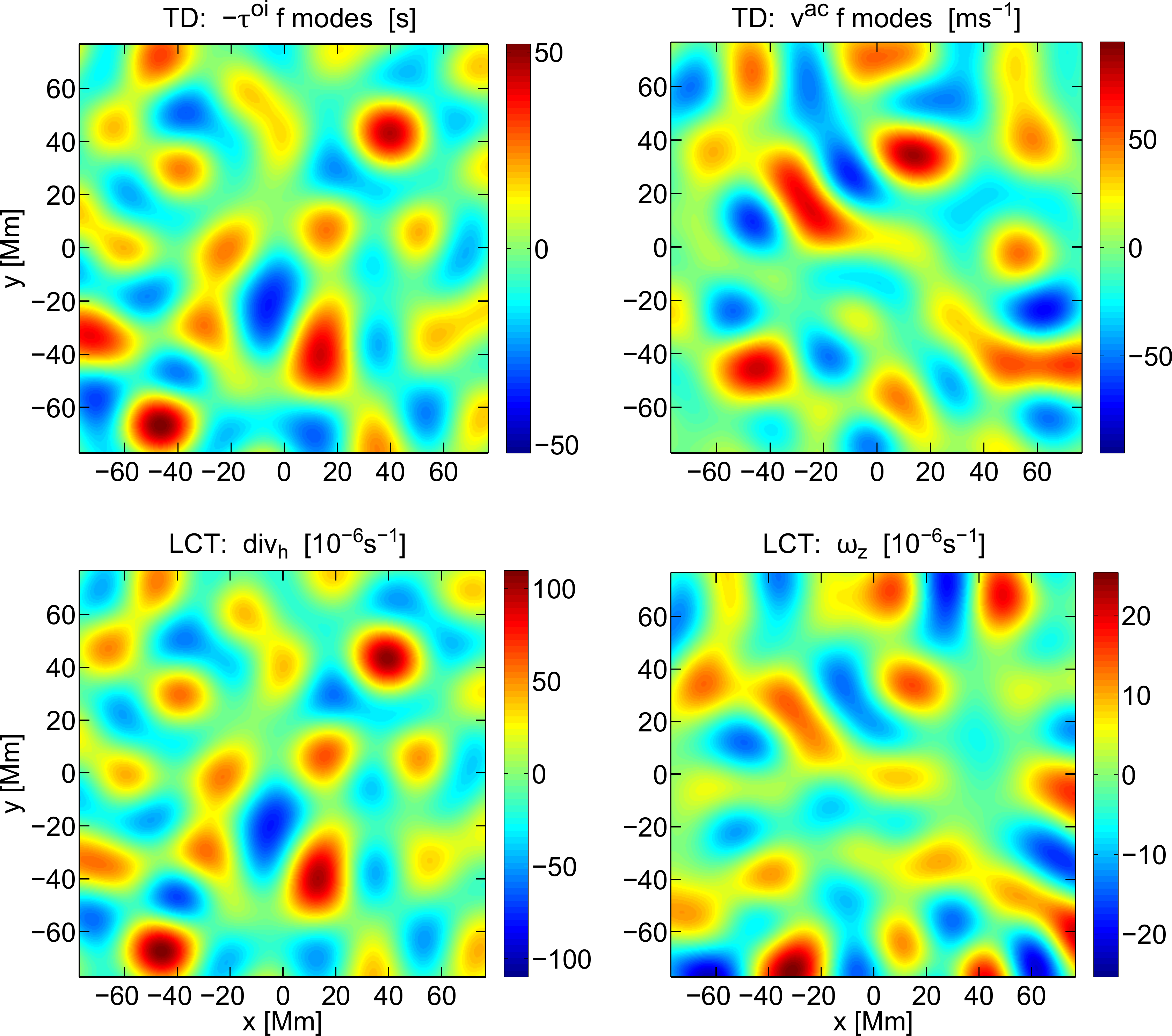}
   \caption{Comparison of TD and LCT maps at the equator. All maps are based on 8~h of HMI images (intensity and Doppler velocity) taken on 1 May 2010 and have been band-pass-filtered around $kR_\odot = 100$. The colorbar limits are set to the maximum absolute value of the corresponding map and symmetrized around zero. \textbf{Left column:} Divergence-sensitive travel times $\tau^\text{oi}$ for f modes as well as LCT horizontal divergence $\text{div}_h$. \textbf{Right column:} Circulation velocities $v^\text{ac}$ for TD as well as LCT vertical vorticity $\omega_z$. The TD $\tau^\text{ac}$ maps have been converted into velocity maps by pointwise multiplication with a constant factor $-5.62$~m~s$^{-2}$ (f modes) and $-11.1$~m~s$^{-2}$ (p$_1$ modes), see Appendix~\ref{chap_conversion} for details.}
\label{fig_lct-vs-tt}
    \end{figure*}

For comparing maps of horizontal divergence and vertical vorticity, one point to consider is the different spatial sampling for TD and LCT maps. To correct for this, we interpolate the velocity maps derived from LCT onto the (finer) travel-time grid. In order to compare the maps on different spatial scales, we apply different band-pass filters to the individual maps in Fourier space. The individual filters are centered around $kR_\odot$ values of 50 through 400 in steps of 50. Each filter is one in a plateau region of width 50, centered around these values. Adjacent to both sides of the plateau are raised cosine flanks that make the filter smoothly reach zero within a $kR_\odot$ range of 50. Additionally, we employ a high-pass filter for $kR_\odot > 400$. From all maps, we subtract the respective mean map over all 336 datasets prior to filtering.

Example 8~h maps for $\tau^\text{oi}$ and $v^\text{ac}$ from f-mode travel times as well as $\text{div}_h$ and $\omega_z$ from LCT are depicted in Fig.~\ref{fig_lct-vs-tt}. The maps are filtered around $kR_\odot = 100$. Note that for the sake of an easier comparison, we plotted $-\tau^\text{oi}$ rather than $\tau^\text{oi}$. For the flow divergence, all three maps are highly correlated. The average correlation coefficients over all 336 maps are $0.96$ between LCT $\text{div}_h$ and $-\tau^\text{oi}$ for f modes and $0.92$ between LCT $\text{div}_h$ and $-\tau^\text{oi}$ for p$_1$ modes.

In the case of flow vorticity, the agreement of the LCT and TD maps is weaker than for the divergence. The average correlation coefficient over all 336 maps is $0.68$ between LCT $\omega_z$ and f-mode $v^\text{ac}$ and $0.51$ between LCT $\omega_z$ and p$_1$-mode $v^\text{ac}$ (not shown). For comparing LCT $v^\text{ac}$ instead of $\omega_z$ with TD $v^\text{ac}$, the correlation coefficients are noticably higher ($0.75$ for f modes and $0.57$ for p$_1$ modes). The flow magnitudes are roughly comparable.

   \begin{table}[h!]
     \caption{Correlation between LCT maps and TD travel-time maps derived from HMI intensity images and Dopplergrams.} 
\label{table_correlation}      
\centering                          
\begin{tabular}{c c|c c c}        
\hline\hline                 
 & & \multicolumn{3}{c}{Correlation coeff. between LCT and TD} \\
Modes & $kR_\odot$ & LCT div$_h$ & LCT $\omega_z$ & LCT $v^\text{ac}$ \\
(TD) & & TD $-\tau^\text{oi}$ & TD $-\tau^\text{ac}$ & TD $-\tau^\text{ac}$ \\
\hline              
f    & 50  & $0.93$ & $0.70$ & $0.77$ \\
     & 100 & $0.96$ & $0.68$ & $0.75$ \\
     & 150 & $0.96$ & $0.63$ & $0.68$ \\
     & 200 & $0.94$ & $0.53$ & $0.57$ \\
     & 250 & $0.89$ & $0.31$ & $0.30$ \\
     & 300 & $0.78$ & $-0.01$ & $0.14$ \\
     & 350 & $0.58$ & $-0.09$ & $0.23$ \\
     & 400 & $0.31$ & $-0.06$ & $0.23$ \\
     & >400 & $0.02$ & $0.00$ & $0.03$ \\ \hline
p$_1$    & 50  & $0.90$ & $0.53$ & $0.59$ \\
         & 100 & $0.92$ & $0.51$ & $0.57$ \\
         & 150 & $0.89$ & $0.44$ & $0.50$ \\
         & 200 & $0.83$ & $0.33$ & $0.38$ \\
         & 250 & $0.66$ & $0.13$ & $0.19$ \\
         & 300 & $0.36$ & $-0.06$ & $0.11$ \\
         & 350 & $0.15$ & $-0.05$ & $0.11$ \\
         & 400 & $0.04$ & $-0.02$ & $0.08$ \\
         & >400 & $0.00$ & $0.00$ & $0.01$ \\ 
\hline  
\end{tabular}
   \end{table}

Table~\ref{table_correlation} shows the correlation coefficients between LCT and TD averaged over all datasets for all filters and including p$_1$ modes. The error in the correlation coefficients is less than 0.01. Note that the edges (12~Mm) were removed from the maps before the correlation coefficients were computed. For the flow divergence, the correlation coefficients are almost constantly high for smaller $kR_\odot$ values. In the range $kR_\odot = 300 - 400$, the correlation coefficient between LCT $\text{div}_h$ and $-\tau^\text{oi}$ for f modes rapidly decreases from $0.78$ to $0.31$. For LCT and p$_1$ modes, the correlation coefficient decreases from $0.83$ to $0.15$ from $kR_\odot = 200 - 350$. For the high-pass filters, the LCT and TD maps are completely uncorrelated.

In the case of vorticity, the correlation decreases rapidly for both f and p$_1$ modes at $kR_\odot = 200$. Again, the LCT and TD maps are uncorrelated for large $kR_\odot$. The correlation coefficients for LCT $v^\text{ac}$ are significantly higher than for LCT $\omega_z$.

   \begin{figure*}
  \centering
\includegraphics[width=\hsize]{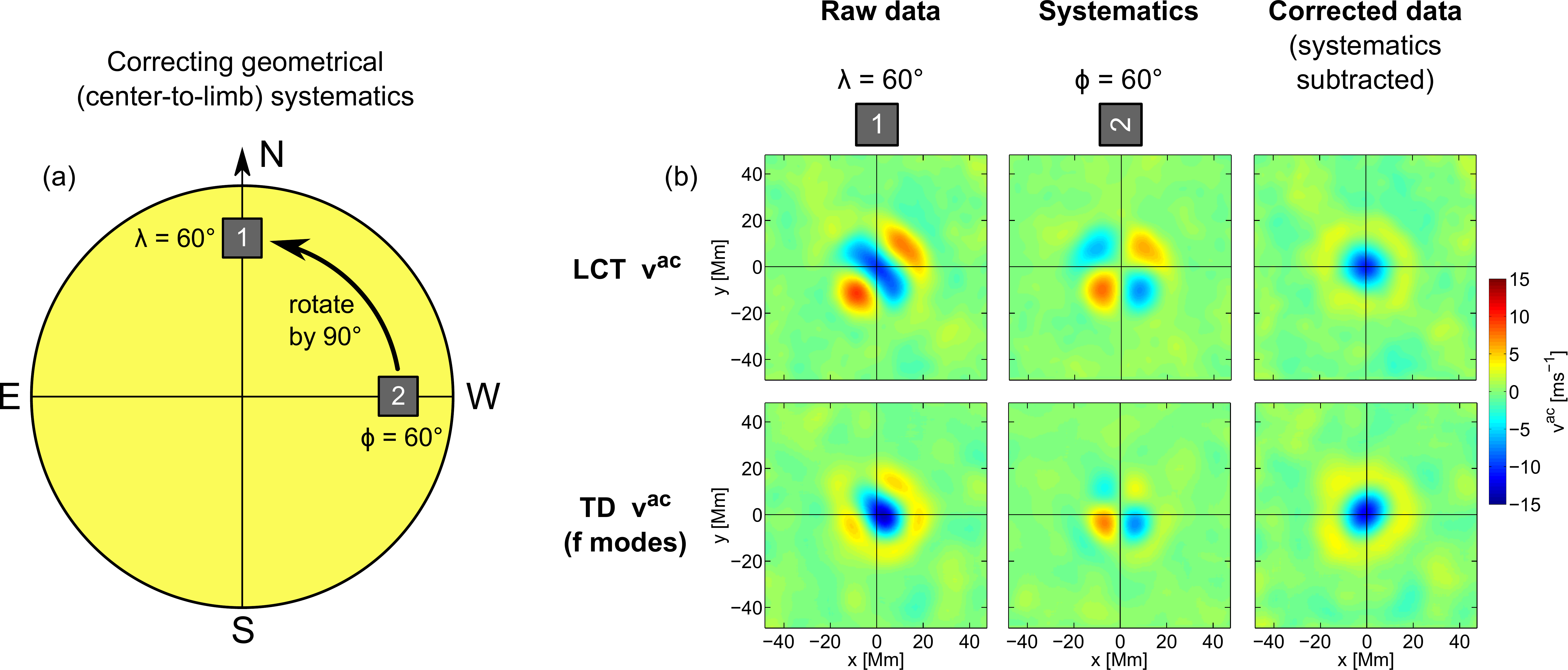}
\caption{Correction for geometrical center-to-limb systematics in maps of circulation velocity $v^\text{ac}$ for LCT and TD using HMI observations. This example shows the correction of $v^\text{ac}$ maps at $60\degr$ latitude. \textbf{a)} Patch 1 (raw data) is located at $60\degr$ latitude at the central meridian. Patch 2 (systematics data) is situated at the equator, $60\degr$ west of the central meridian. We correct the raw data by subtracting patch 2 from patch 1 (after rotating patch 2 by $90\degr$ anti-clockwise). Note that we also use measurements of the systematics from the east for the correction (rotated $90\degr$ clockwise). This is omitted in the sketch for clarity. \textbf{b)} Maps of $v^\text{ac}$ for LCT and f-mode TD. The three columns of panels show (from left to right) the measured raw data at $60\degr$ latitude, the measured systematics from $60\degr$ west and east (rotated and averaged), and the corrected data.}
\label{fig_center-to-limb-correction}
    \end{figure*}

The dependence of the correlation coefficients on spatial scale conceptually agrees well with the power spectra in Fig.~\ref{fig_power}. There is a high correlation on large scales where the observed TD travel-time power clearly exceeds the power of the TD noise model. On the other hand, the very low correlation on smaller scales reflects that the power of TD observations and noise model are almost equal.

Qualitatively, the correlation coefficients are comparable with the value $0.89$ from \citet{derosa_2000} who obtained travel-time and LCT velocity maps from SOHO/MDI Dopplergrams and smoothed the divergence maps by convolving with a Gaussian with FWHM $6.2$~Mm.


\section{Net vortical flows in the average supergranule}
The major goal of this paper is to spatially resolve the vorticity of the average supergranule at different solar latitudes. In the following, we describe the averaging process and show average divergence and vorticity maps.

\subsection{Obtaining maps of the average supergranule}
To construct the average supergranule, we started by identifying the location of supergranule outflows and inflows in f-mode $\tau^\text{oi}$ maps from HMI and MDI. We smoothed the maps by removing power for $kR_\odot>300$ and applied an image segmentation algorithm \citep[cf.][]{hirzberger_2008}. The coordinates of the individual supergranules were used to align maps of various data products. For each identified position, we translated a copy of the map to move the corresponding supergranule to the map center. These translated maps were then averaged. At each latitude, we averaged over roughly 3\,000 supergranules in total for HMI (1500 supergranules for MDI). Supergranules closer than 8~Mm to the map edges were discarded.

We produced maps for the average supergranule outflow and inflow from $\tau^\text{oi}$ and $\tau^\text{ac}$ travel-time maps as well as LCT $v_x$ and $v_y$ maps. Prior to the averaging process, the LCT maps were spatially interpolated onto the (finer) travel-time grid. For all maps, we subtracted the mean map over all 336 HMI datasets (177 datasets in the case of MDI). This removes signal that does not (or only slowly) change with time, including differential rotation. Additionally, we removed power for $kR_\odot>300$ by applying a low-pass filter in Fourier space.

The resulting average $\tau^\text{ac}$ maps were converted into $v^\text{ac}$ maps. From the LCT $v_x$ and $v_y$ maps for the average supergranule, we computed $\text{div}_h$, $\omega_z$, and $v^\text{ac}$. The Savitzky-Golay filters that we employed to compute the spatial derivatives smooth out step artefacts from the image alignment process, yet preserve the signal down to the resolution limit of the LCT.

We corrected the $v^\text{ac}$ and $\omega_z$ maps for geometrical center-to-limb systematics (unless stated otherwise). We measured these effects using HMI and MDI observations west and east of the disk center, at relative longitudes corresponding to the latitudes of the regular observations. The idea is that any difference (beyond the noise background) between maps at disk center and a location west or east from disk center is due to geometrical center-to-limb systematics. Such systematics only depend on the distance to the disk center. Therefore, our raw measurements of $v^\text{ac}$ and $\omega_z$ that we obtained north and south of the equator should be affected by the systematics in the same way as measurements west and east of the disk center. We corrected the raw data by subtracting the $v^\text{ac}$ and $\omega_z$ maps west and east of the disk center. This approach is analogous to \citet{zhao_2013} who used the method to correct measurements of the meridional circulation.
Figure~\ref{fig_center-to-limb-correction} illustrates the correction process for $v^\text{ac}$ maps at $60\degr$ latitude. The correction is particularly important for LCT at high latitudes. We note that the measured center-to-limb systematics at lower latitudes (up to $40\degr$ north and south) are much weaker and only lead to a mild correction of the $v^\text{ac}$ and $\omega_z$ maps.
A further discussion of the center-to-limb systematics can be found in Appendix~\ref{sect_center-to-limb-systematics}.

\subsection{Latitudinal dependence of the vertical vorticity in outflow regions} \label{sect_lat-dependence}
\begin{figure}
  \centering
 \includegraphics[width=0.985\hsize]{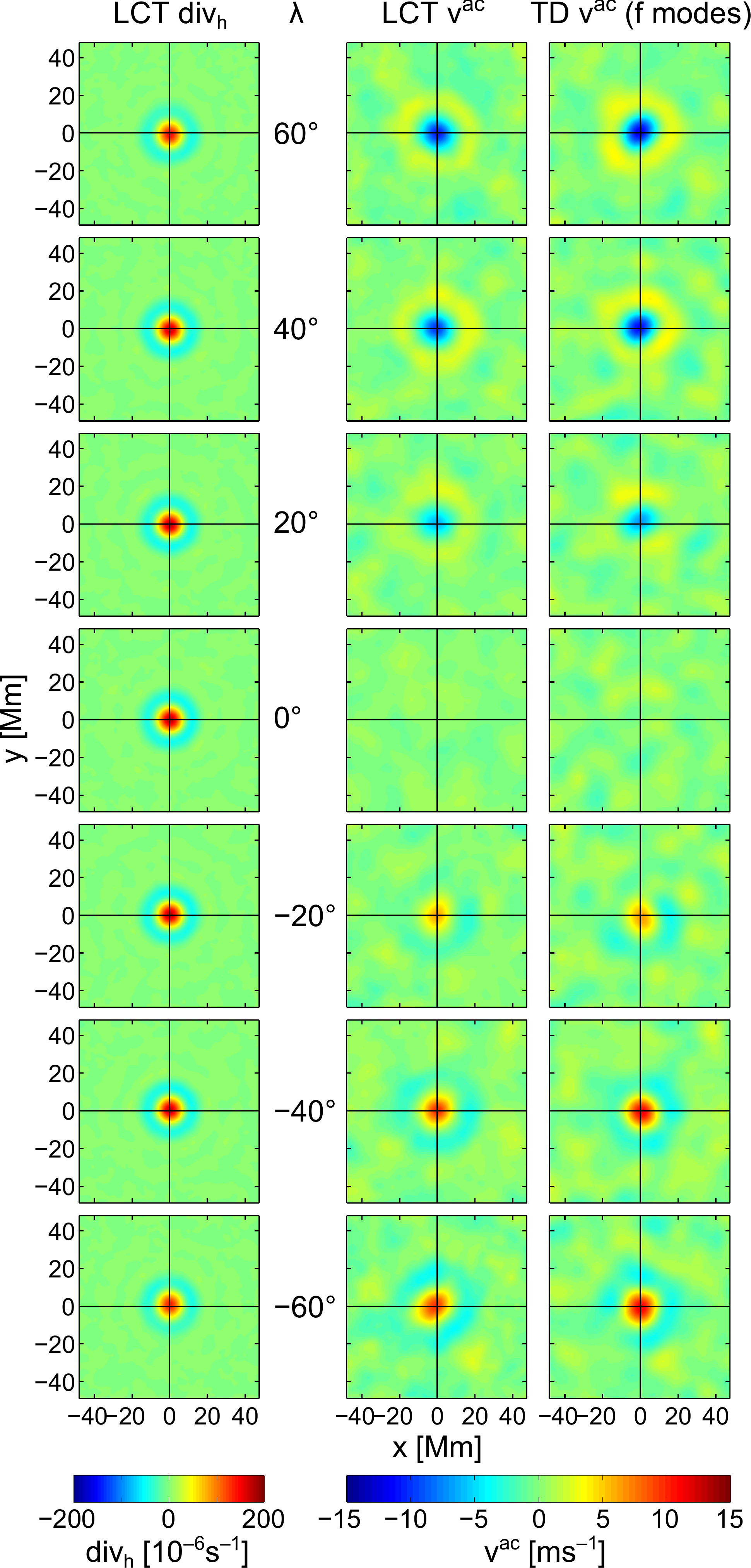}
   \caption{Maps of horizontal divergence $\text{div}_h$ and circulation velocity $v^\text{ac}$ for the average supergranule outflow regions at various solar latitudes (after correction for geometrical center-to-limb systematics). The maps have been derived from HMI intensity images and Dopplergrams. The horizontal divergence $\text{div}_h$ has been computed from LCT $v_x$ and $v_y$ horizontal velocity maps. The LCT $v^\text{ac}$ maps were obtained by averaging the tangential velocity component $v_t$ over an annulus with radius 10~Mm and half-width 2~Mm to resemble the $\tau^\text{ac}$ measurement geometry (Fig.~\ref{fig_geometry}d). For TD, the $v^\text{ac}$ maps are based on $\tau^\text{ac}$ travel-time maps (Fig.~\ref{fig_geometry}b) that have been computed from Dopplergrams.}
\label{fig_aligned-ccw-outflow}
    \end{figure}

Figure~\ref{fig_aligned-ccw-outflow} shows the circulation velocity $v^\text{ac}$ in the average supergranule outflow region for LCT and f-mode TD for latitudes from $-60\degr$ to $60\degr$, in steps of $20\degr$. For comparison, the left column shows the horizontal divergence div$_h$ from LCT. At all latitudes, there is a peak of positive divergence at the origin. All divergence peaks are surrounded by rings of negative divergence. This reflects that on average every supergranule outflow region is isotropically surrounded by inflow regions. The strength of the divergence peak slightly decreases toward higher latitudes. Furthermore, the divergence peaks are slightly shifted equatorward at high latitudes (by about $0.7~$Mm at $\pm60\degr$). These effects are presumably due to center-to-limb systematics.

The $v^\text{ac}$ maps (center and right columns) show negative peaks (clockwise motion) in the northern hemisphere and positive peaks (anti-clockwise motion) in the southern hemisphere. The peaks are surrounded by rings of opposite sign, as for the divergence maps. There is a remarkable agreement between LCT and TD in both shape and strength of the peak structures.
At the solar equator, no peak and ring structures are visible. Note that the LCT and TD $v^\text{ac}$ maps at the equator are still correlated though. This shows that the ``noise'' background is due to real flows rather than measurement noise that is dependent on the technique.

\begin{figure*}
  \sidecaption
\includegraphics[width=0.35\hsize]{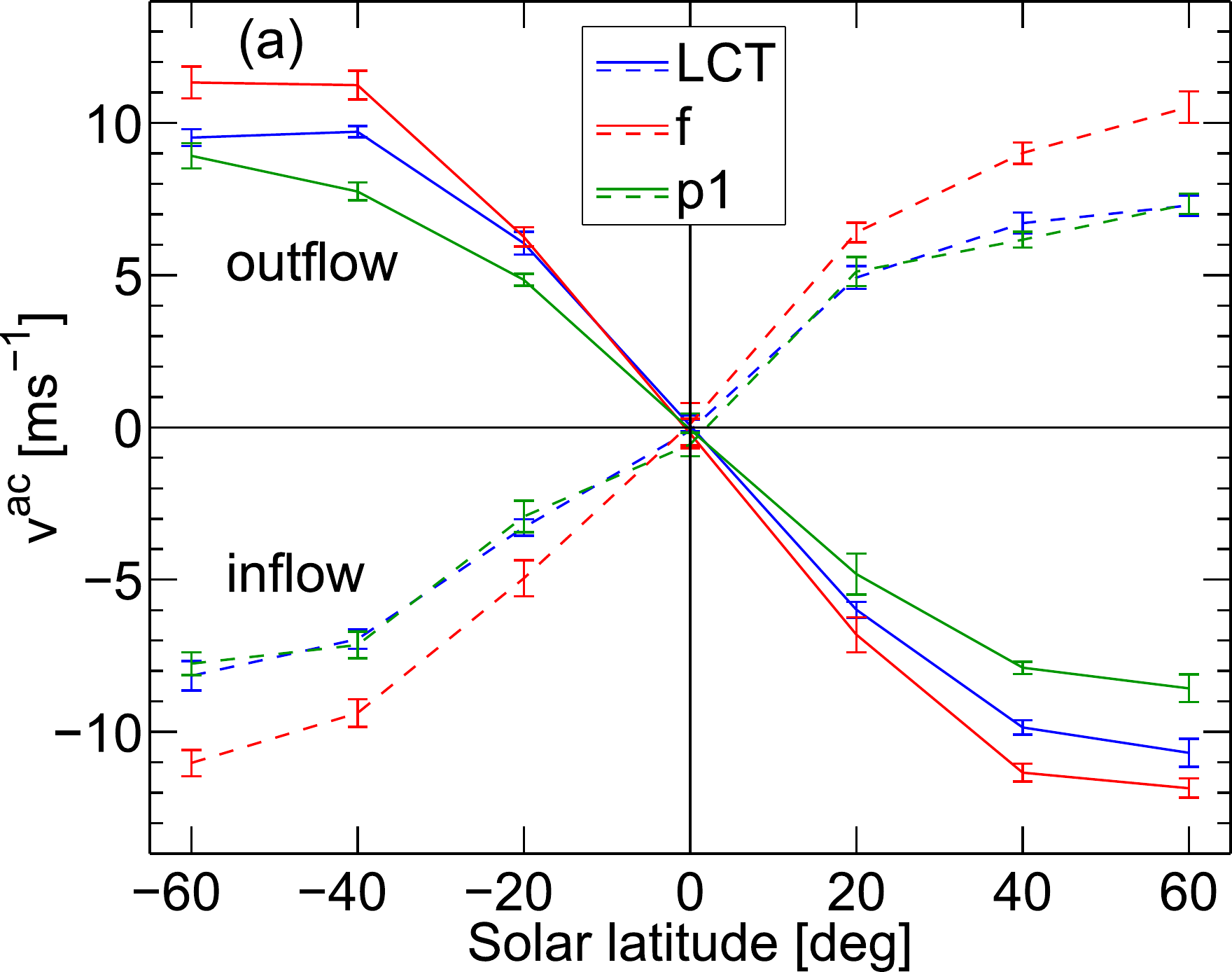}
\includegraphics[width=0.35\hsize]{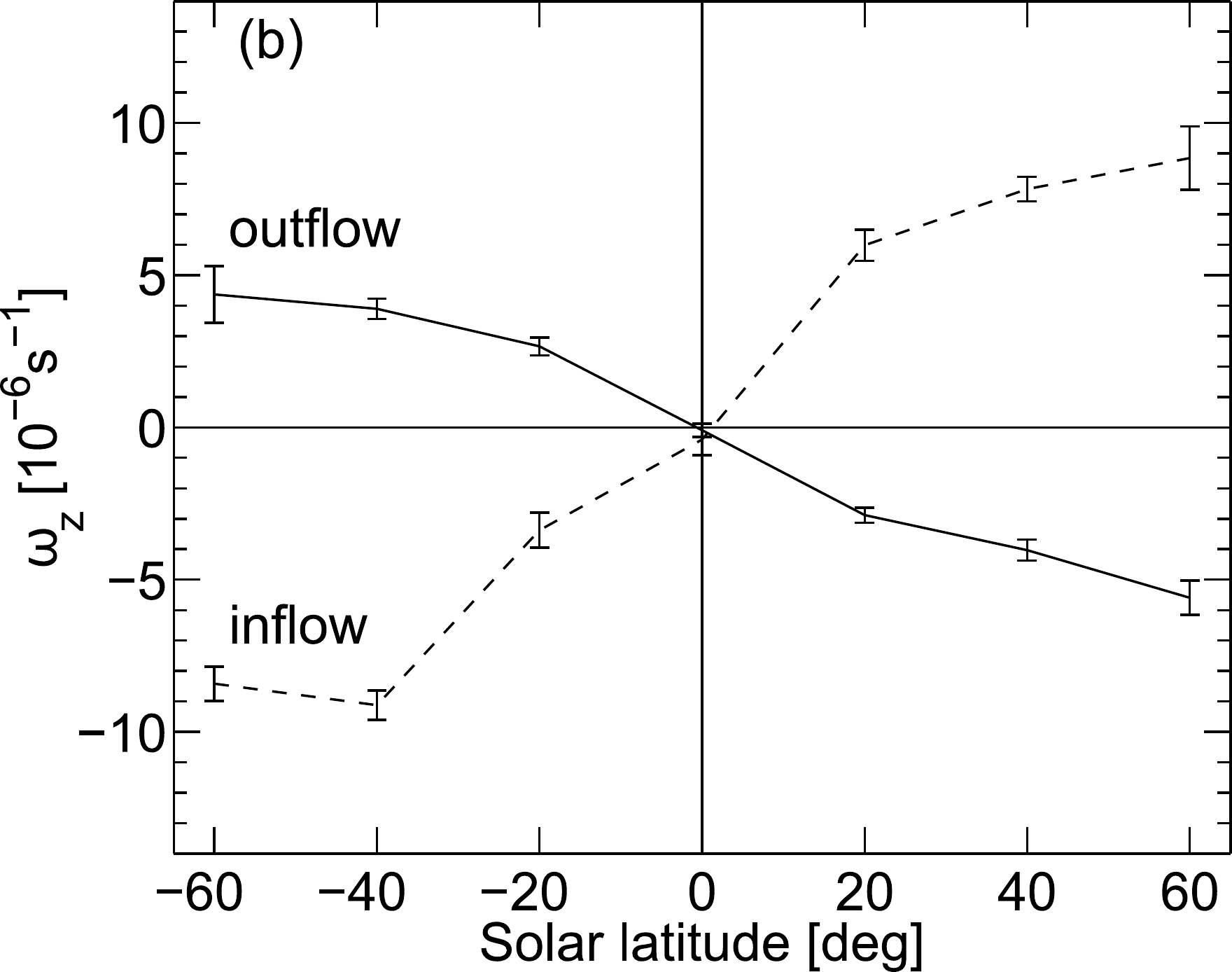}
   \caption{Peak $v^\text{ac}$ and $\omega_z$ values for the average supergranule at different solar latitudes. \textbf{a)} Circulation velocity $v^\text{ac}$ for LCT, f modes and p$_1$ modes. \textbf{b)} Vertical component of flow vorticity $\omega_z$ obtained from LCT. Solid lines are for the average supergranule outflow region, dashed lines for the average supergranule inflow region. At $0\degr$ latitude, the values at the map center are shown instead of the peak values. The errorbars have been computed from dividing the 336 datasets into eight parts and measuring the variance of $v^\text{ac}$ and $\omega_z$ at the peak positions over the eight parts.}
\label{fig_aligned_vs_lat}
    \end{figure*}

To study the latitudinal dependence of the observed and corrected signal in more detail, we plot in Fig.~\ref{fig_aligned_vs_lat}a the peak velocity $v^\text{ac}$ from Fig.~\ref{fig_aligned-ccw-outflow}, including p$_1$-mode TD, as a function of solar latitude (lines). The peak velocity shows an overall decrease from south to north, with a zero-crossing at the equator. The curves are antisymmetric with respect to the origin. The peak velocities have similar values at a given latitude, with f-mode velocities appearing slightly stronger than LCT and p$_1$-mode velocities (in this order). The highest velocities are slightly above 10~m~s$^{-1}$. Figure~\ref{fig_aligned_vs_lat}b shows the peak magnitude in maps of the vertical vorticity $\omega_z$, as measured from LCT. The overall appearance is similar to the circulation velocities $v^\text{ac}$. The highest absolute vorticity value is about $5\times 10^{-6}$~s$^{-1}$.

   \begin{figure*}[h]
  \sidecaption
\includegraphics[width=12.95cm]{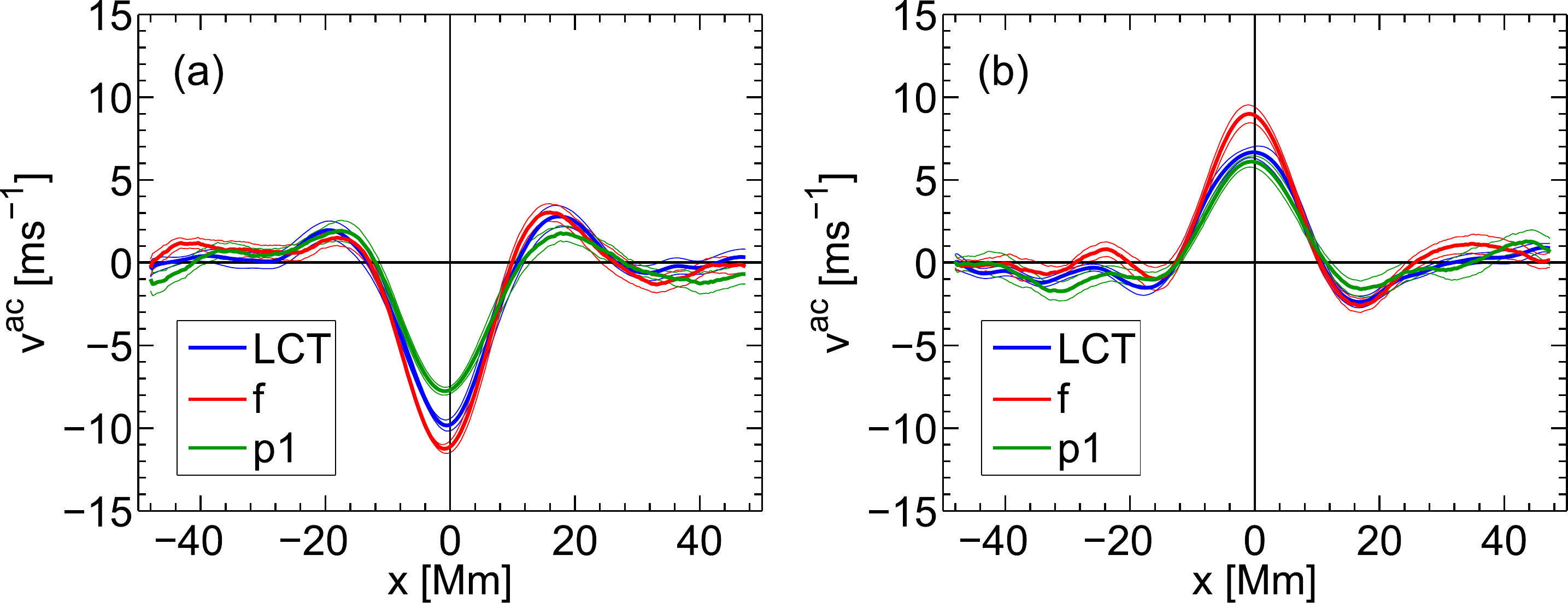}
   \caption{\textbf{a)} Cuts through the maps of the circulation velocity $v^\text{ac}$ for the average supergranule outflow region at $40\degr$ latitude (shown in Fig.~\ref{fig_aligned-ccw-outflow}), at $y=0$. The TD and LCT maps were derived from HMI Dopplergrams and intensity images. The thin lines denote estimates of the variability of the data as obtained from dividing the 336 datasets into eight parts. The $1\sigma$ level is shown. \textbf{b)} As \textbf{a)}, but for the average supergranule inflow region at $40\degr$ latitude.}
\label{fig_aligned_cut}
    \end{figure*}

Figure~\ref{fig_aligned_cut}a shows cuts through $y=0$ for the maps of LCT and TD $v^\text{ac}$ (including p$_1$ modes) at $40\degr$ latitude. We choose this latitude because the S/N in the $v^\text{ac}$ and $\omega_z$ peaks is high compared to other latitudes, whereas the measurements are only mildly affected by center-to-limb systematics. The velocity magnitudes and shapes of the curves are comparable for the three cases. For the LCT and f-mode curves, an asymmetry in the west-east direction is visible. This means that the ring structures surrounding the peaks in the $v^\text{ac}$ maps are stronger in the west than in the east. The FWHM is about 13~Mm in all cases.
The peaks are very slightly shifted eastwards. However, this east shift does not appear to be a general feature at all latitudes. Mostly, the shifts are consistent with random fluctuations. Partly, the shifts might also be due to other effects, for instance an incomplete removal of center-to-limb systematics.

For comparison, the FWHM of the $\tau^\text{oi}$ peak structure is about 13~Mm for p$_1$ modes, compared to about 11~Mm for f-mode $\tau^\text{oi}$. The horizontal divergence $\text{div}_h$ from LCT at $40\degr$ latitude peaks at about $170\times 10^{-6}$~s$^{-1}$ with a FWHM of about 10~Mm.

From the $v^\text{ac}$ peak velocities, we can estimate the average vorticity $\langle\omega_z\rangle_A$ over the circular area $A$ of radius $R=10$~Mm that is enclosed by the $\tau^\text{ac}$ measurement contour (see Fig.~\ref{fig_geometry}). The average vorticity is given by
\begin{equation}
 \langle\omega_z\rangle_A = \frac{\Gamma}{A} \approx \frac{2v^\text{ac}}{R} \ ,   \label{eq_circulation}
\end{equation}
where $\Gamma$ is the flow circulation along the $\tau^\text{ac}$ measurement contour that we approximated with $\Gamma \approx 2\pi R v^\text{ac}$. By taking the $v^\text{ac}$ peak values, we obtain $\langle\omega_z\rangle_A \approx -2.4\times10^{-6}~$s$^{-1}$ for the f modes, $\langle\omega_z\rangle_A \approx -1.6\times10^{-6}~$s$^{-1}$ for the p$_1$ modes, and $\langle\omega_z\rangle_A \approx -2.0\times10^{-6}~$s$^{-1}$ for LCT. Thus the average vorticity in the circular region is roughly half the peak vorticity at $40\degr$ latitude.

\subsection{Inflow regions} \label{sect_inflows}
\begin{figure}[h!]
  \centering
 \includegraphics[width=0.985\hsize]{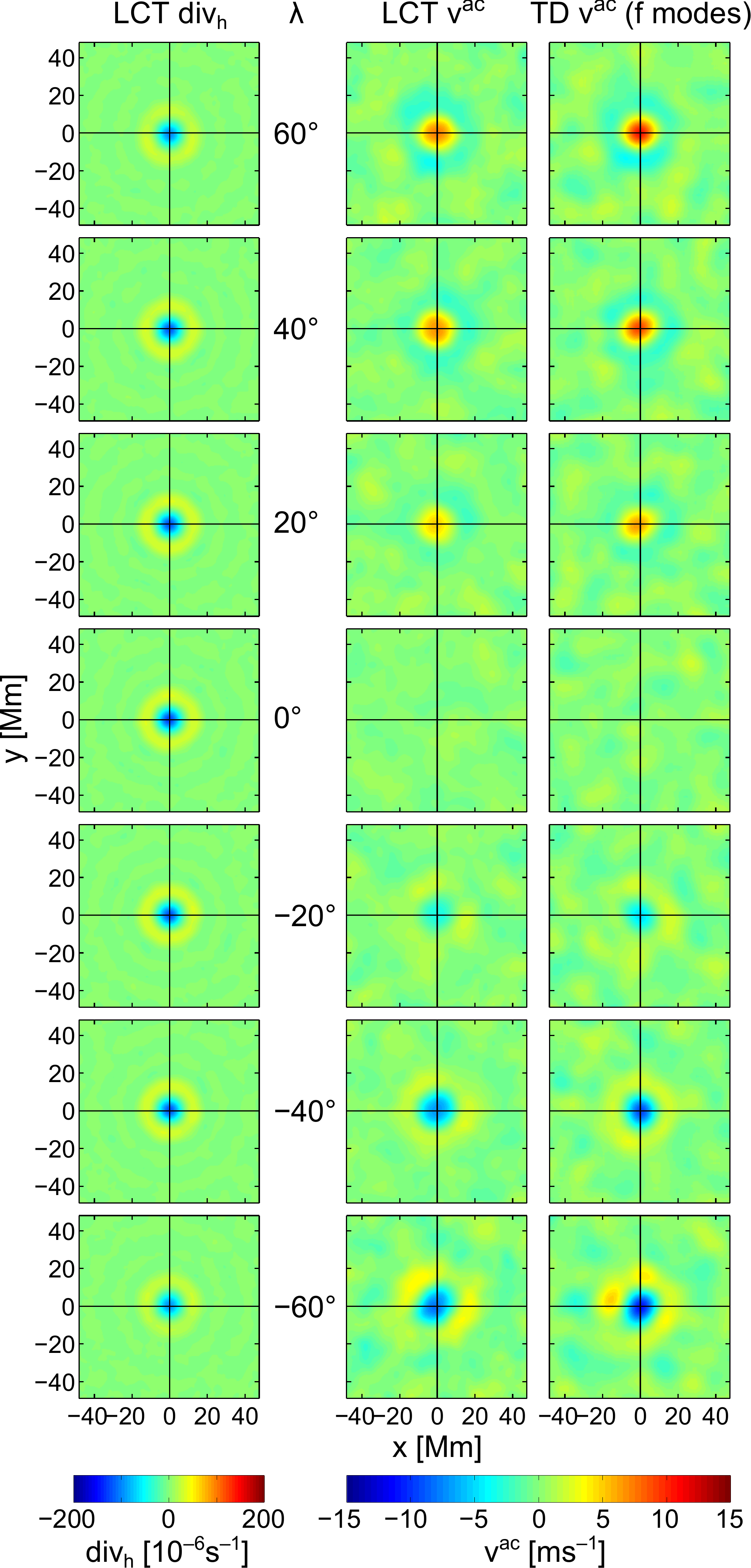}
   \caption{Same as Fig.~\ref{fig_aligned-ccw-outflow}, but for the average supergranule inflow regions.}
\label{fig_aligned-ccw-inflow}
    \end{figure}

So far, we have discussed vortical flows around supergranule outflow centers. It is interesting though to compare the magnitude and profile of these flows with the average inflow regions, which have a different geometrical structure (connected network instead of isolated cells). Analogue to Fig.~\ref{fig_aligned-ccw-outflow} for the outflows, Fig.~\ref{fig_aligned-ccw-inflow} shows maps of $\text{div}_h$ and $v^\text{ac}$ around the average supergranule inflow center. As for the outflows, the $v^\text{ac}$ maps from TD and LCT agree very well at all analyzed latitudes. The peaks in the $v^\text{ac}$ maps have the opposite sign compared to the outflows. This indicates that flows are preferentially in the clockwise (anti-clockwise) direction in the average supergranular outflow region and anti-clockwise (clockwise) in the average inflow region in the northern (southern) hemisphere. Cuts through $y=0$ of the $v^\text{ac}$ maps at $40\degr$ latitude are shown in Fig.~\ref{fig_aligned_cut}b. The $v^\text{ac}$ curves have the same shape as the corresponding curves for the average outflow center (with a FWHM of 14 to 16~Mm) but the peak flow magnitude is reduced and the sign is switched. As for the outflows, the ring structures are stronger on the west side than on the east side.

The horizontal flow divergence div$_h$ in the average inflow is similar to the average outflow (about the same FWHM) but with reversed signs and reduced magnitude. The peak divergence is about $-120\times 10^{-6}$~s$^{-1}$ at $40\degr$ latitude with a FWHM of about 10~Mm. As for the outflows, there is a systematic decrease in peak magnitude and a slight equatorward shift of the div$_h$ peak at high latitudes.

The latitude dependence of the $v^\text{ac}$ peak values for the average supergranule inflow region (dashed lines in Fig.~\ref{fig_aligned_vs_lat}a) is almost mirror-symmetric to the outflow regions. The values are slightly smaller though compared to the average outflow, with a ratio inflow/outflow of $-0.87\pm0.03$ for the f modes, $-0.85\pm0.06$ for the p$_1$ modes, and $-0.72\pm0.05$ for the LCT $v^\text{ac}$. In the case of $\omega_z$ (Fig.~\ref{fig_aligned_vs_lat}b), on the other hand, the ratio between the average inflow and outflow center is $-1.8\pm0.2$.

From the peak values of $v^\text{ac}$, we can estimate the average vorticity $\langle\omega_z\rangle_A$ over the circular area $A$ of radius $R=10$~Mm in the same way as for the outflow regions. We obtain $\langle\omega_z\rangle_A \approx 1.8\times10^{-6}~$s$^{-1}$ for the f modes, $\langle\omega_z\rangle_A \approx 1.2\times10^{-6}~$s$^{-1}$ for the p$_1$ modes, and $\langle\omega_z\rangle_A \approx 1.4\times10^{-6}~$s$^{-1}$ for LCT. The peak vorticity at $40\degr$ latitude is therefore larger by a factor of about five compared to the vorticity averaged over the circular area.

\subsection{Dependence of the vertical vorticity on horizontal divergence}
The detection of net tangential flows in the average supergranule raises the question of how much the magnitudes of $v^\text{ac}$ and $\omega_z$ depend on the selection of supergranules. As a test, we sort the identified supergranules at $40\degr$ latitude from HMI with respect to their divergence strength, as measured by the peak f-mode $\tau^\text{oi}$ of each supergranule. The sorted supergranules are assigned to four bins, which each contain roughly the same number of supergranules. The boundaries of the bins for f-mode $\tau^\text{oi}$ are about $-96.9$, $-53.8$, $-42.1$, $-31.5$, and $-16.0$~s for the outflows and $67.5$, $38.3$, $32.1$, $26.0$, and $11.1$~s for the inflows. Note that a simple scatter plot would be very noisy because the $v^\text{ac}$ and $\omega_z$ maps are dominated by turbulence.

   \begin{figure*}
\sidecaption
\includegraphics[width=0.33\hsize]{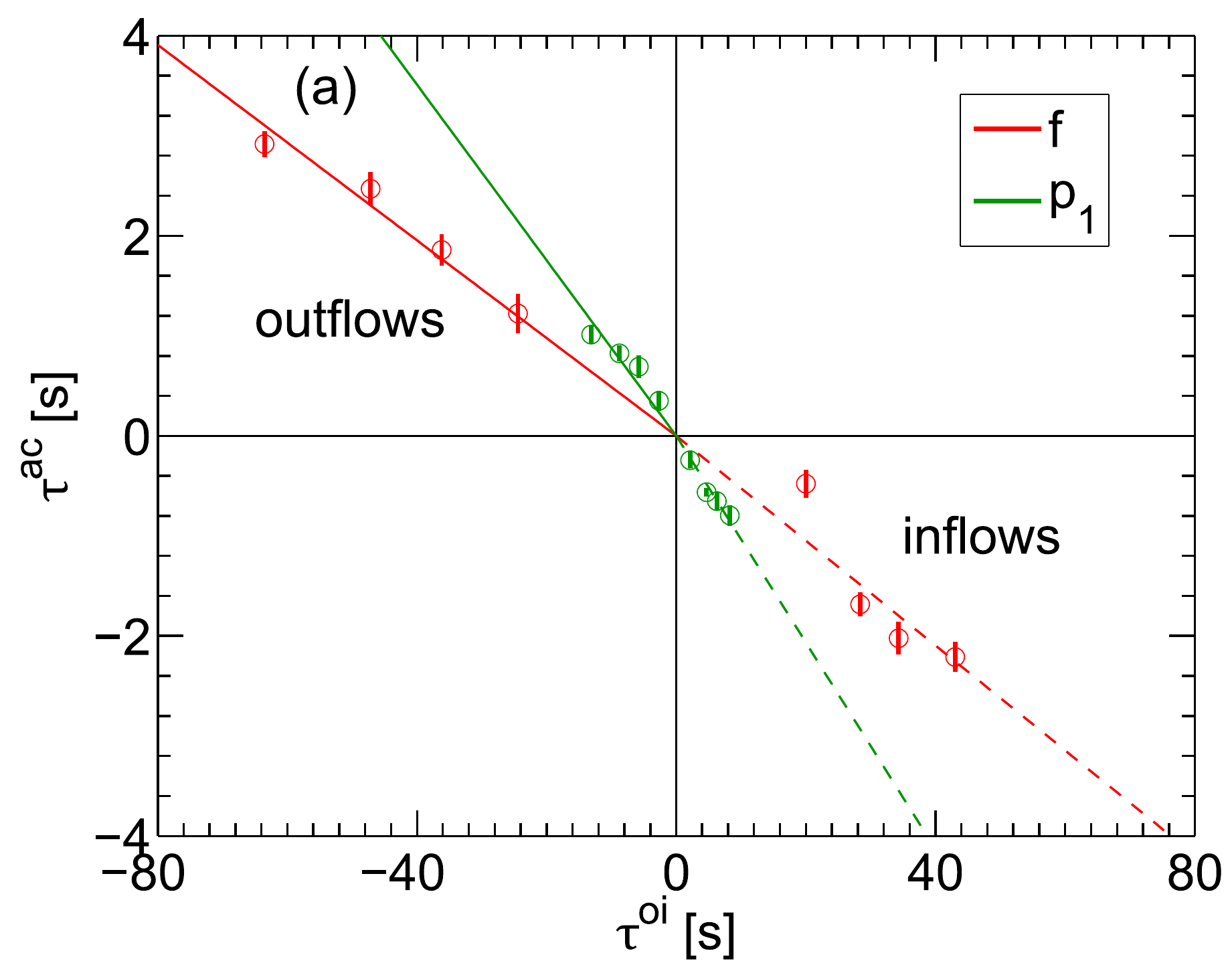}
\includegraphics[width=0.33\hsize]{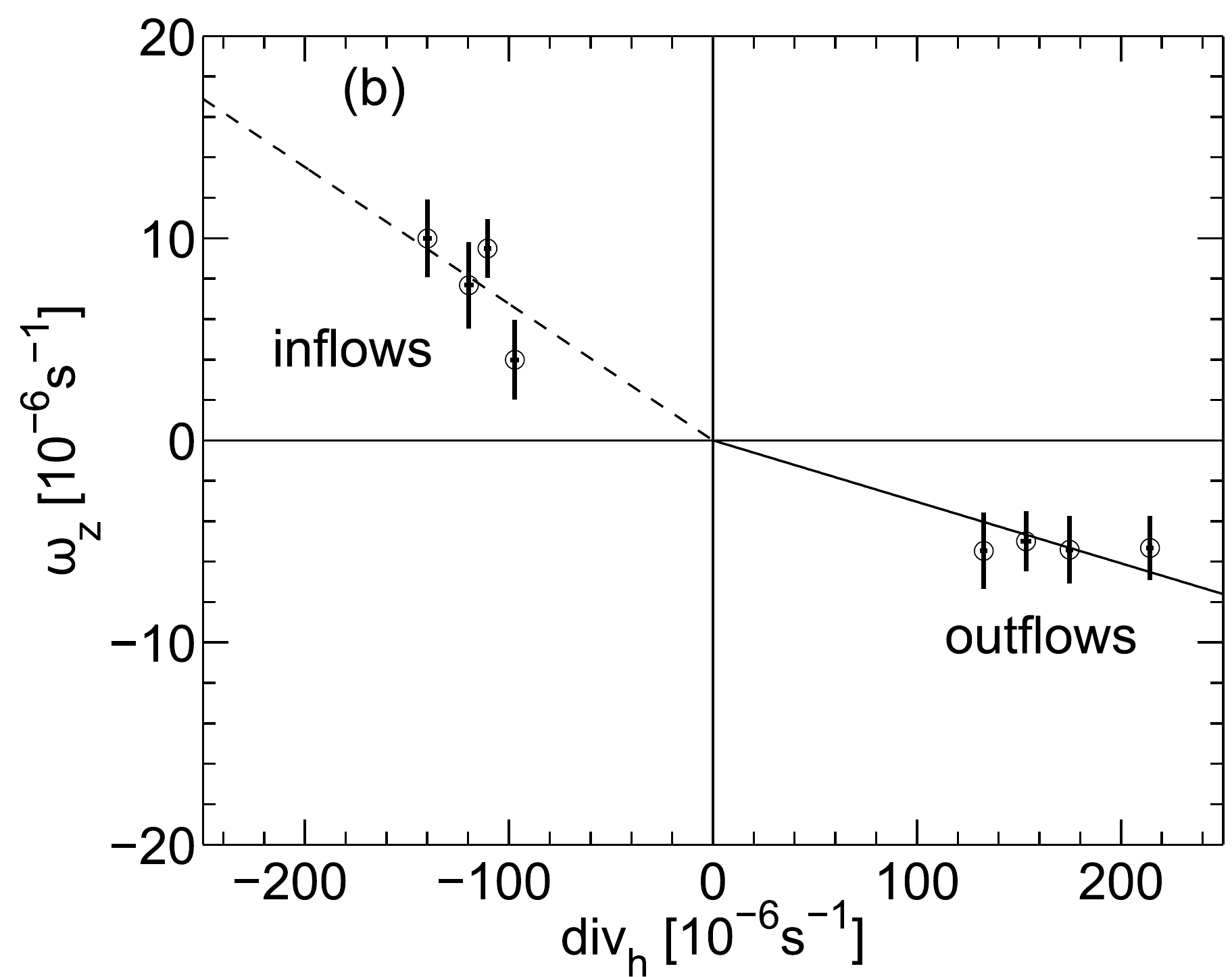}
   \caption{Vorticity as a function of divergence for the average supergranule at $40\degr$ latitude from HMI data. \textbf{a)} Vorticity travel times $\tau^\text{ac}$ versus divergence travel times $\tau^\text{oi}$ for TD f modes and p$_1$ modes. The upper-left quadrant shows the values for the average outflows, the lower-right quadrant for the average inflows. The solid and the dashed lines show least-squares fits of a linear function through the origin for outflows and inflows, respectively. The errorbars were obtained from dividing the 336 datasets into eight parts. \textbf{b)} As a) but the peak $\omega_z$ versus the peak $\text{div}_h$ from LCT is shown. Note that the quadrants depicting outflows and inflows are flipped compared to the travel times in  \textbf{a)}.}
\label{fig_SG_strength}
    \end{figure*}

For each bin, we computed the peak TD $\tau^\text{oi}$ and $\tau^\text{ac}$ as well as LCT $\text{div}_h$ and $\omega_z$ in the same way as for all identified supergranules that we discussed in the previous sections, but without the correction for center-to-limb systematics. In Fig.~\ref{fig_SG_strength}a, we plot the peak $\tau^\text{ac}$ as a function of the peak $\tau^\text{oi}$ from f modes and p$_1$ modes both for outflows and inflows. The magnitude of $\tau^\text{ac}$ clearly increases with $\tau^\text{oi}$. The ratio $\tau^\text{ac}/\tau^\text{oi}$ is roughly constant. Only the f-mode bin for the weakest inflows deviates substantially from this behaviour.
Figure~\ref{fig_SG_strength}b shows the peak $\omega_z$ versus the peak $\text{div}_h$ from LCT. In this case, the relationship is less clear, considering the large vertical errorbars. A constant ratio $\omega_z/\text{div}_h$ is, at least, by eye consistent with the measurements. However, for outflows $\omega_z$ might also be constant. Note that the fit lines for the travel times in Fig.~\ref{fig_SG_strength}a have almost the same slopes for outflows and inflows, whereas in the case of LCT $\text{div}_h$ and $\omega_z$ the slope for the inflows is much steeper than for the outflows. This is consistent with Fig.~\ref{fig_aligned_vs_lat}, where $\omega_z$ was shown to be twice as strong in the inflows as in the outflows, whereas the velocities $v^\text{ac}$ are of similar magnitude (not just for TD, but also for LCT). As discussed in Sects.~\ref{sect_lat-dependence} and \ref{sect_inflows}, the velocities $v^\text{ac}$ do not directly measure the vorticity at a given position but rather a spatial average.

In general, we can conclude that a selection bias in favour of stronger or weaker supergranules probably does not affect the measured ratio of vertical vorticity to horizontal divergence.

\subsection{Comparison of SDO/HMI and SOHO/MDI}
While the results for the average supergranule have been obtained using different methods (TD and LCT) and image types (Dopplergrams and intensity images), they are all based on the same instrument -- HMI. It is thus useful to compare the HMI results to $v^\text{ac}$ maps that have been measured from independent MDI data. Since MDI cannot sufficiently resolve granules at higher latitudes to successfully perform LCT, however, we only discuss TD.

In contrast to HMI, the correction for geometric center-to-limb systematics is not sufficient for MDI. For example, for f-mode TD at $40\degr$ latitude, the central peak structure appears elongated (see Fig.~\ref{fig_MDI_outflow} in the appendix). Nevertheless, the $v^\text{ac}$ values at the origin are remarkably similar for HMI and MDI. At the average outflow, we measure $(-11.1\pm0.4)$~m~s$^{-1}$ (HMI) versus $(-10.1\pm0.8)$~m~s$^{-1}$ (MDI) for f modes and $(-7.7\pm0.3)$~m~s$^{-1}$ compared to $(-6.5\pm0.8)$~m~s$^{-1}$ for p$_1$ modes.

For inflows, the MDI $v^\text{ac}$ maps compare to HMI in the same manner, with MDI being slightly weaker than HMI. The flow magnitudes for HMI and MDI at the origin after correction are $(8.9\pm0.4)$~m~s$^{-1}$ versus $(7.2\pm1.0)$~m~s$^{-1}$ for f modes and $(6.1\pm0.3)$~m~s$^{-1}$ compared to $(4.2\pm0.5)$~m~s$^{-1}$ for p$_1$ modes.
Note that the noise background is stronger for MDI. This is, however, not surprising, since only about half the number of Dopplergrams (compared to HMI) have been used to produce these maps.

   \begin{figure}[h!]
  \centering
\includegraphics[width=0.8\hsize]{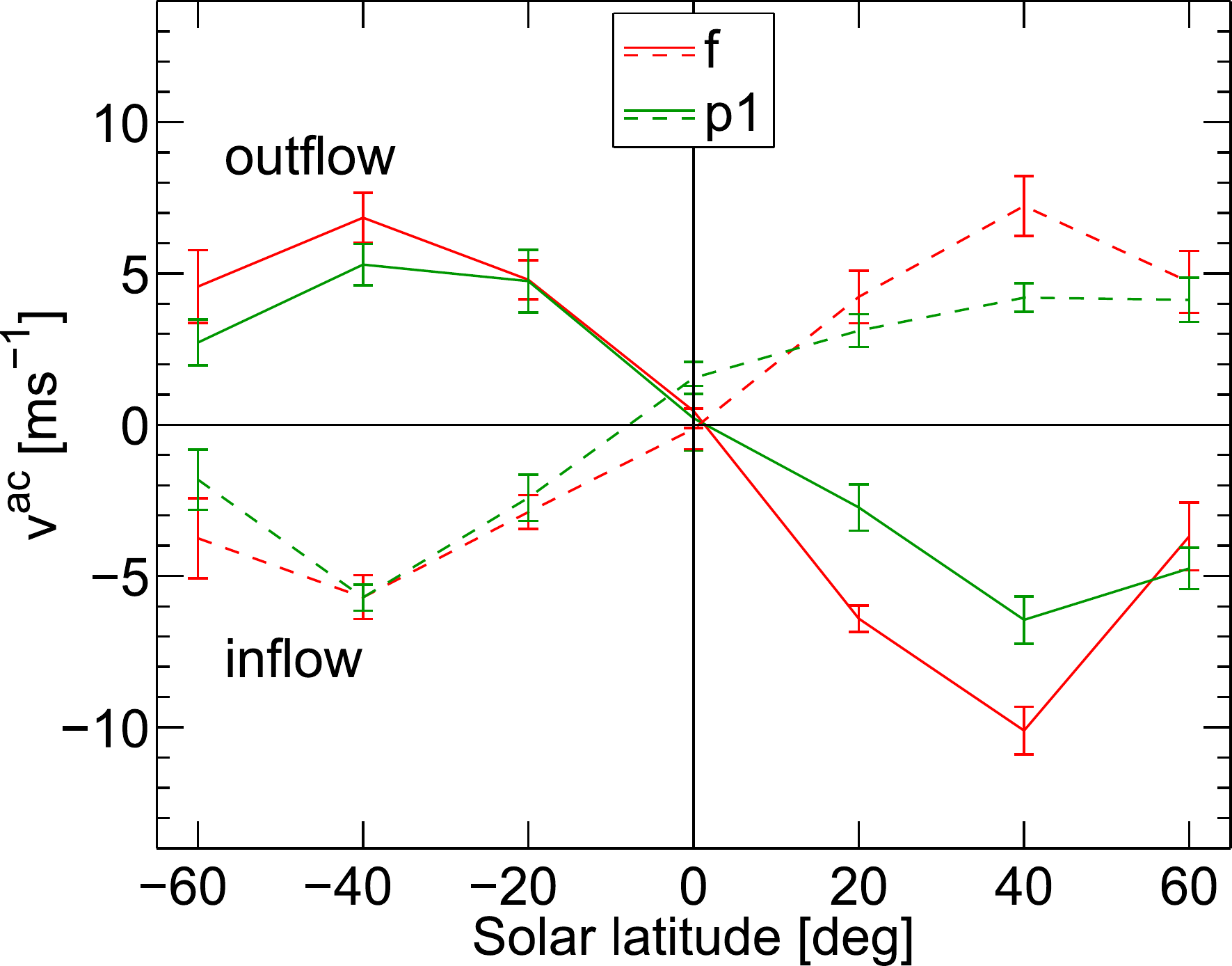}
   \caption{Velocities $v^\text{ac}$ for the average supergranule measured at the origin at different solar latitudes from MDI TD (f modes and p$_1$ modes). Solid lines are for the average supergranule outflow region, dashed lines for the average supergranule inflow region. The errorbars have been computed from dividing the 177 datasets into eight parts.}
\label{fig_aligned_vs_lat_MDI}
    \end{figure}

The latitude dependence of $v^\text{ac}$ at the origin for MDI is qualitatively comparable with HMI (see Fig.~\ref{fig_aligned_vs_lat_MDI}). We measure a zero-crossing and sign change of $v^\text{ac}$ at the equator, both for the average supergranule outflow and inflow regions. However, the $v^\text{ac}$ magnitudes are systematically smaller for MDI. This difference increases farther away from the equator. It is especially dramatic for f modes at $\pm60\degr$ latitude. Whereas $v^\text{ac}$ reaches values between 10 and 12~m~s$^{-1}$ at these latitudes in HMI, for MDI the velocity magnitudes are below 5~m~s$^{-1}$. This is probably connected to the lower spatial resolution of MDI, which results in a larger impact of geometrical foreshortening effects at high latitudes compared to HMI. 

While MDI clearly performs much worse than HMI, the agreement with HMI at the origin gives reason to believe that MDI $v^\text{ac}$ measurements are still useful. This would be especially interesting for long-term studies of the solar cycle dependence since continuous data reaching back to 1996 could be used.


\section{Differences between outflow and inflow regions}

   \begin{figure*}[h!]
  \sidecaption
\includegraphics[width=0.35\hsize]{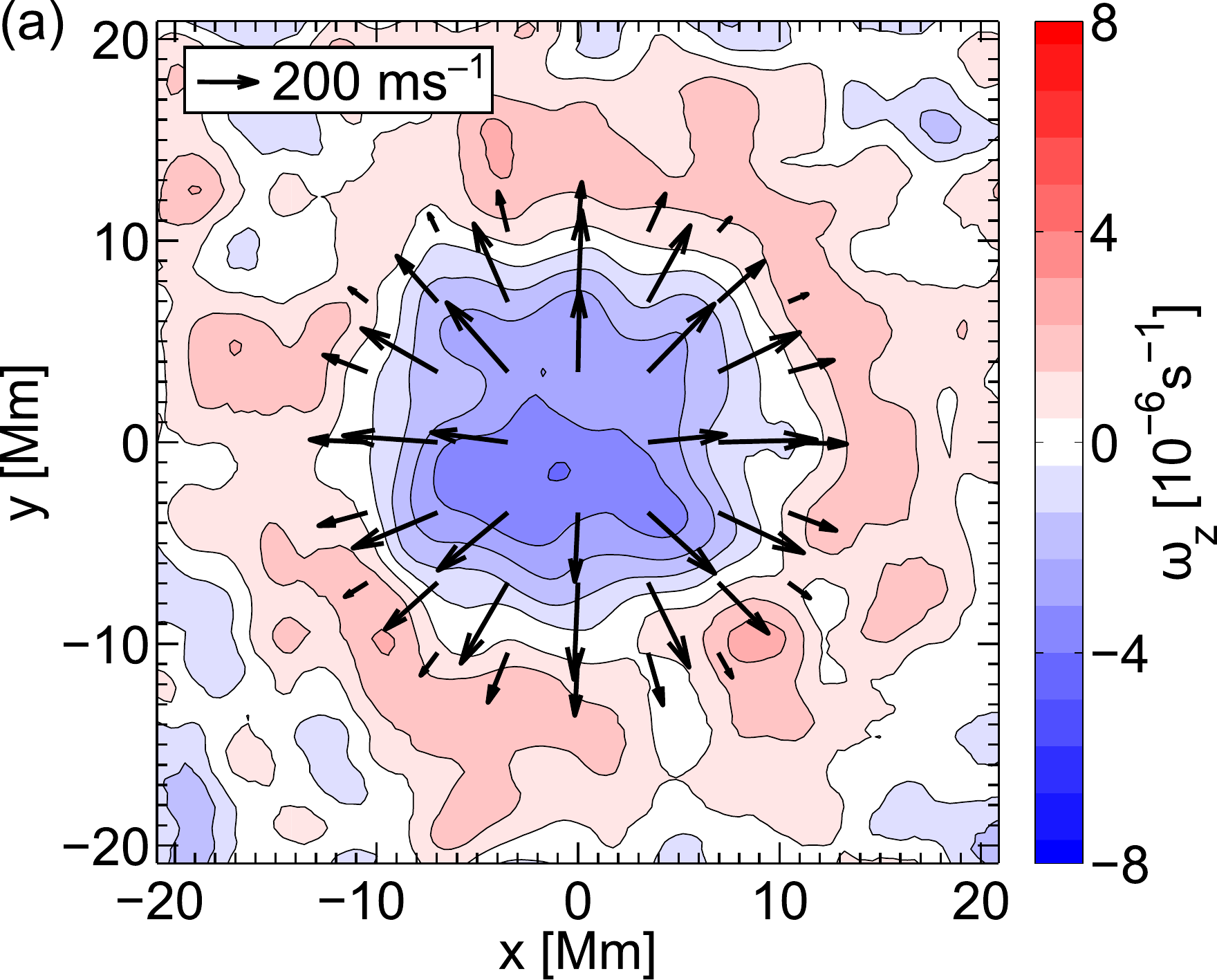}
\includegraphics[width=0.35\hsize]{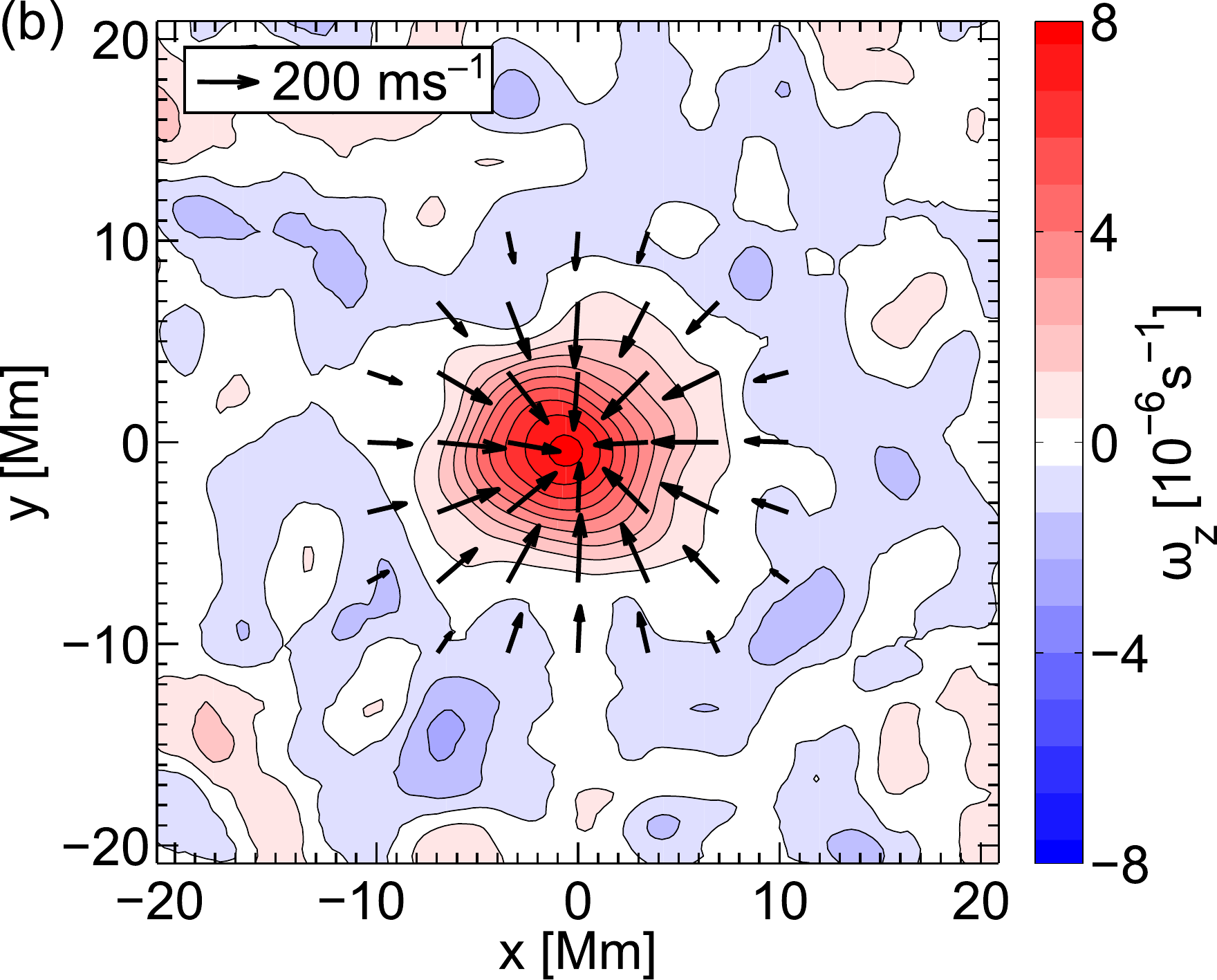}
   \caption{LCT horizontal velocity (black arrows) and vertical vorticity $\omega_z$ (filled contours) for the average supergranule at solar latitude $40\degr$. \textbf{a)} Average outflow region. \textbf{b)} Average inflow region. Arrows for velocity magnitudes less than 60~m~s$^{-1}$ are omitted.}
\label{fig_summary}
    \end{figure*}

The differences between the average supergranule outflow and inflow regions as measured from LCT in HMI data are summarized in Fig.~\ref{fig_summary}. The arrows show the horizontal velocity magnitudes and directions at 40$\degr$ latitude. The flows are dominated by the radial velocity component. For direct comparison, the filled contours give the vertical vorticity $\omega_z$ of the flows. In the average outflow region (Fig.~\ref{fig_summary}a), the vorticity shows a broad plateau region (FWHM about 13~Mm). The region of negative vorticity is surrounded by a ring of positive vorticity with a diameter of about 30~Mm.

In contrast, the vorticity in the average inflow region (Fig.~\ref{fig_summary}b) falls off rapidly from its narrow center (FWHM 8~Mm). We note that the FWHM of the vorticity peak is smaller than for the divergence peak (about 10~Mm) but still larger than the FWHM of the LCT correlation measurements (roughly 3~Mm). The peak vorticity magnitude is about twice the value of the outflow region (about $8\times 10^{-6}$~s$^{-1}$ anti-clockwise compared to $4\times 10^{-6}$~s$^{-1}$ clockwise). Like for the average outflow region, the central vorticity structure in the inflow region is surrounded by a ring of vorticity with opposite sign. The vorticity magnitude in the ring appears to be smaller than in Fig.~\ref{fig_summary}a.

These differences in the vortex structures of outflow and inflow regions are visible at all latitudes (except at the equator, where we measure no net vorticity). The FWHM of the peak structures as well as the ratio of the peak vorticities (between outflow and inflow regions) are constant over the entire observed latitude range. Such differences do not appear in maps of the horizontal divergence $\text{div}_h$ (the FWHM is about 10~Mm in both outflow and inflow regions).

Differences in the vorticity strength between regions of divergent and convergent flows have also been reported by other authors who studied the statistics of vortices in solar convection. \citet{wang_1995} found, on granular scales, the root mean square of $\omega_z$ to be slightly higher in inflow regions. \citet{poetzi_2007} observed that vortices are strongly connected to sinks at mesogranular scales. Concentration of fluid vorticity in inflows has also been found in simulations of solar convection \citep[e.g.,][on granulation scale]{stein_1998}. However, the authors did not find any preferred sign of $\omega_z$.
The increased vorticity strength in inflows might be a manifestation of the ``bathtub effect'' \citep{nordlund_1985}. In that scenario, initially weak vorticity becomes amplified in inflows due to angular momentum conservation. In the downflows that are associated with the horizontal inflows because of mass conservation, the vortex diameter is reduced since the density rapidly increases with depth. This further enhances the vorticity.

\section{Radial and tangential velocities versus radial distance}
\begin{figure*}
  \sidecaption
\includegraphics[width=0.35\hsize]{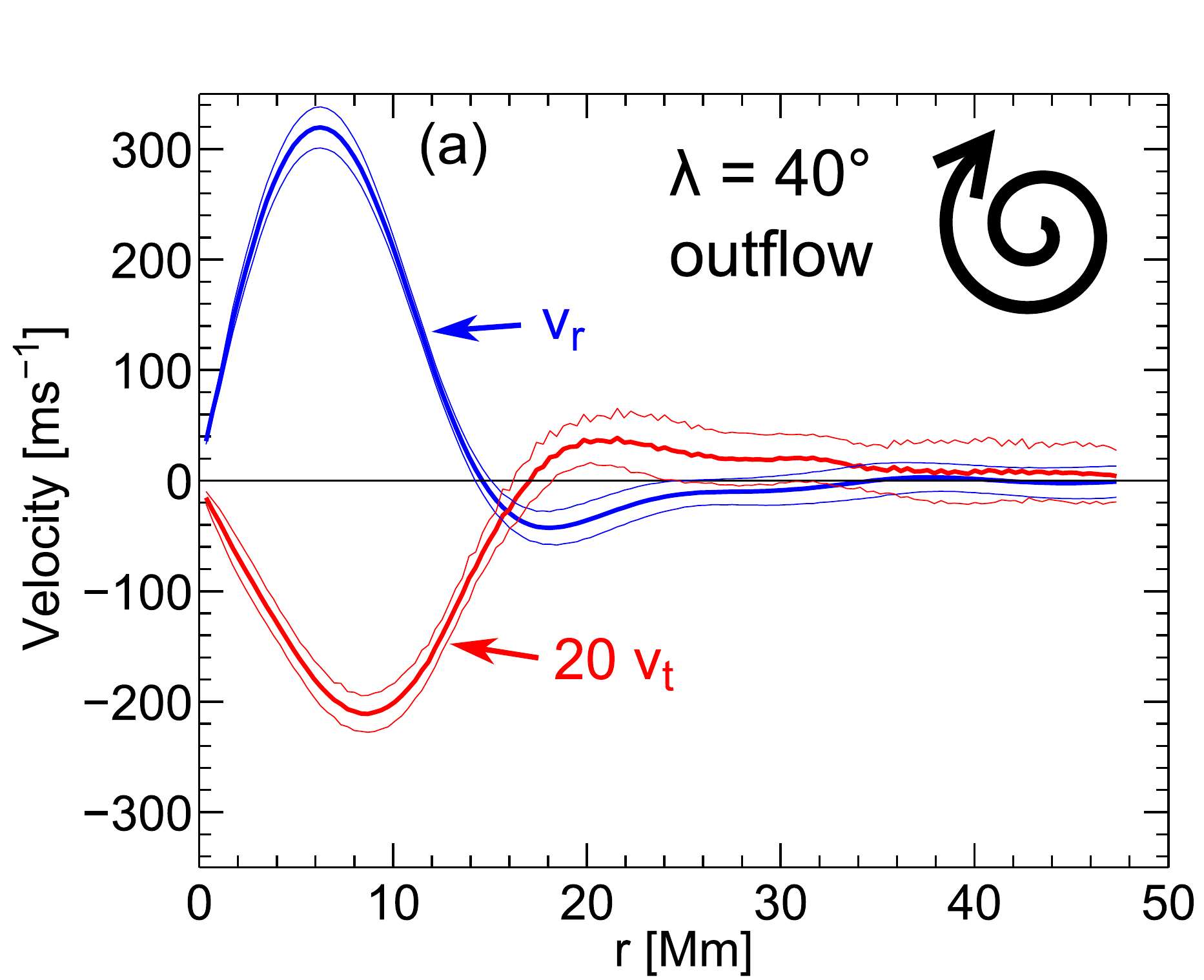}
\includegraphics[width=0.35\hsize]{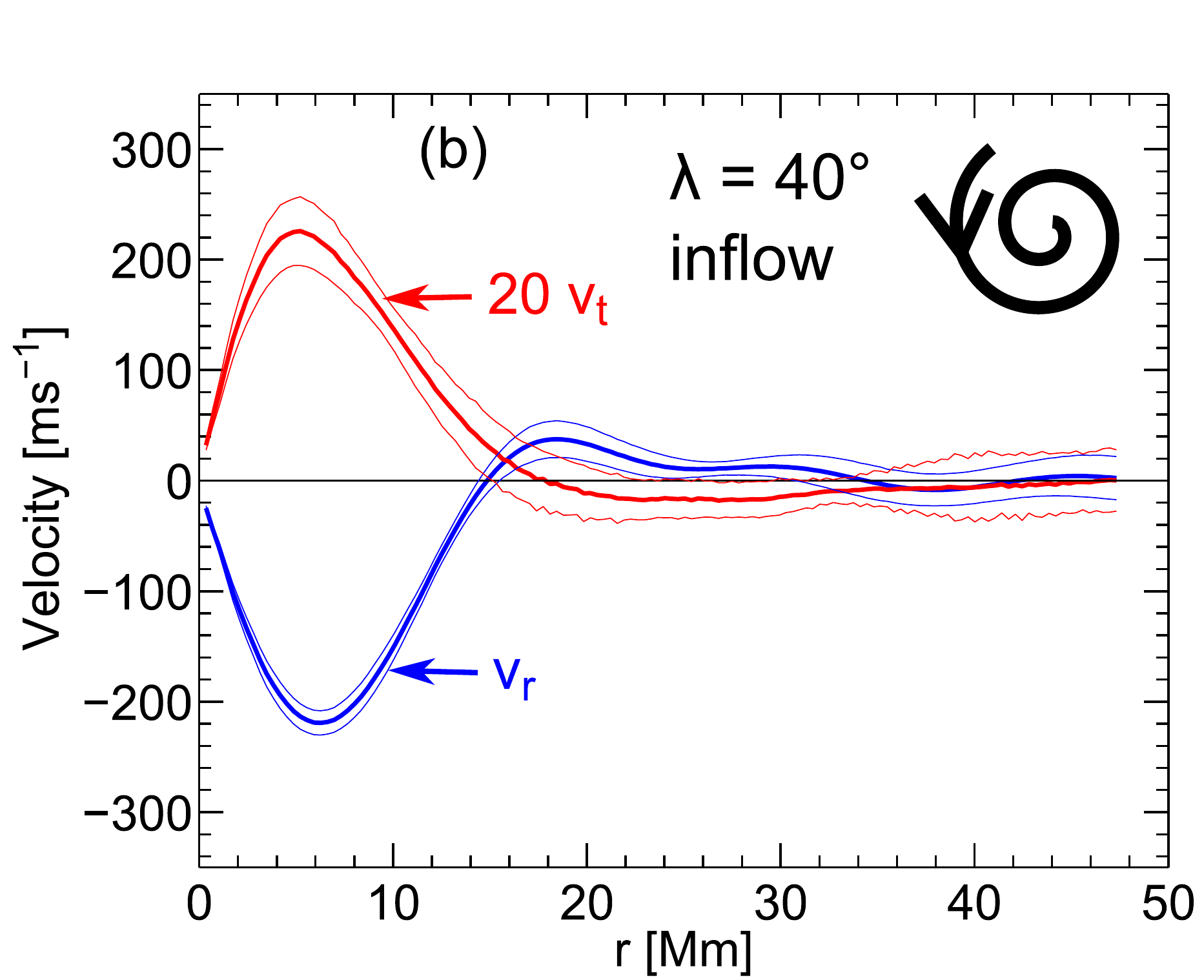}
\caption{Azimuthally averaged horizontal velocities around the average supergranule outflow and inflow centers at solar latitude $40\degr$. The measurements were obtained from LCT of granules in HMI intensity images. \textbf{a)} Horizontal velocities $v_r$ and $v_t$ around the average supergranule outflow center. The thin lines mark an estimate of the variability of the data as measured from dividing the 336 datasets into eight parts. For $v_r$, the $10\sigma$ level is shown, for $v_t$ the $3\sigma$ level. \textbf{b)} As \textbf{a)}, but around the average supergranule inflow center.}
\label{fig_aligned_azi}
    \end{figure*}

Let us now look in greater detail at the isotropic part of the horizontal flow profile of the average supergranule. Figs.~\ref{fig_aligned_azi}a and b show the azimuthal averages of $v_r$ and $v_t$ around both the average supergranule outflow and inflow centers as a function of horizontal distance $r$ to the outflow/inflow center at $40\degr$ latitude.
In both cases, the magnitude of $v_r$ increases from the outflow/inflow center until it reaches a peak velocity (which we call $v_r^\text{max}$) of slightly more than 300~m~s$^{-1}$ and $-200$~m~s$^{-1}$, respectively at $r = 7$~Mm. The flow magnitudes then decrease and $v_r$ switches sign at a distance of about 14~Mm, marking the edge of the average inflow/outflow region. In general, the $v_r$ curves for outflow and inflow regions are similar except for the difference in flow magnitude.

The tangential velocity $v_t$, on the other hand, exhibits similar peak velocities for outflow and inflow regions (both $|v_t^\text{max}| \approx 12~$m~s$^{-1}$) but has opposite signs and reaches these peaks at different distances. The peak magnitude $v_t^\text{max}$ is about 26 times smaller than $v_r^\text{max}$ in the outflow region and 18 times smaller in the inflow region. In the outflow region, the peak is located at $r = 9$~Mm, whereas it lies at $r = 5$~Mm around the average inflow center. Despite the different peak locations, $v_t$ switches sign at a distance of about 17~Mm both around outflow and inflow centers.

The different peak locations of $v_t$ possibly explain why the magnitude ratio of $v^\text{ac}$ between the average supergranule inflow and outflow region is smaller than one. The $v^\text{ac}$ measurements are especially sensitive to $v_t$ at $r = 10$~Mm (the annulus radius). At this distance, we have $v_t = 10$~m~s$^{-1}$ around outflow centers, but $v_t = -7$~m~s$^{-1}$ around inflow centers, yielding a factor of $-0.7$. This agrees well with the ratio of the slopes for LCT $v^\text{ac}$ in Fig.~\ref{fig_aligned_vs_lat} that we discussed in Sect.~\ref{sect_inflows}.

Measuring the peak values of $v_r$ and $v_t$ at all latitudes except the equator leads to the following approximate relations:
\begin{align}
 v_t^\text{max} &= (-0.059\pm0.001) \frac{\Omega(\lambda)\sin\lambda}{\Omega_0} v_r^\text{max}  & \text{for outflows,}  \label{eq_vtmax_outflow} \\
 v_t^\text{max} &= (-0.089\pm0.002) \frac{\Omega(\lambda)\sin\lambda}{\Omega_0} v_r^\text{max}  & \text{for inflows,}  \label{eq_vtmax_inflow}
\end{align}
where we used the differential rotation model from \citet{snodgrass_1984} to compute $\Omega(\lambda)$, and $\Omega_0$ denotes the rotation rate at the equator. The coefficients $b_\text{out} := -0.059\pm0.001$ and $b_\text{in} := -0.089\pm0.002$ are remarkably constant over the whole latitude range from $-60\degr$ to $60\degr$, although $v_r^\text{max}$ decreases from the equator ($335~$m~s$^{-1}$ for outflows and $-237~$m~s$^{-1}$ for inflows) toward high latitudes (e.g., $272~$m~s$^{-1}$ for outflows and $-188~$m~s$^{-1}$ for inflows at $60\degr$ north). The same trend is observed for measuring $v_r^\text{max}$ west and east off disk center (e.g., $272~$m~s$^{-1}$ for outflows and $-187~$m~s$^{-1}$ for inflows at $60\degr$ west), suggesting that the decrease is a systematic center-to-limb effect. Since $b_\text{out}$ and $b_\text{in}$ are not affected, $v_t^\text{max}$ is likely to suffer from the same systematic decrease as $v_r^\text{max}$.

We can use Eqs.~(\ref{eq_vtmax_outflow}) and (\ref{eq_vtmax_inflow}) to predict $v_t^\text{max}$ for supergranules in the polar regions. Assuming $b_\text{out}$ and $b_\text{in}$ are independent of latitude even beyond $\lambda = \pm60\degr$ and employing $v_r^\text{max}$ from the equator where center-to-limb effects are small, the average supergranule at the north pole should rotate with $v_t^\text{max} = (-20~$m~s$^{-1})\times \Omega(90\degr)/\Omega_0$ for outflows and $v_t^\text{max} = (21~$m~s$^{-1})\times \Omega(90\degr)/\Omega_0$ for inflows. At the south pole, merely the sign of $v_t^\text{max}$ should change.

Various authors have proposed models to describe vortex flows \citep[e.g.,][]{taylor_1918,veronis_1959,simon_1997}, introducing the (turbulent) kinematic viscosity $\eta$ as a parameter that influences the tangential velocity component $v_t(r)$.
In the \citeauthor{veronis_1959} model, the tangential flow is given by
\begin{equation}
 v_t = - \frac{2l^2}{\pi^2 (4+l^2/d^2)} \frac{\Omega(\lambda) \sin\lambda}{\eta} v_r,
\end{equation}
where $l$ and $d$ are the horizontal size and depth of the supergranule, respectively. This relationship is consistent with the measurements in Eqs.~(\ref{eq_vtmax_outflow}) and (\ref{eq_vtmax_inflow}).
Note, though, that the \citeauthor{veronis_1959} convection model does not include turbulence (beyond $\eta$) or stratification.

\citet{taylor_1918} presented a simple model that describes the decay of a narrow isolated vortex with $v_r = 0$ due to fluid viscosity. In this case, the tangential velocity component is 
\begin{equation}
v_t(r) = \frac{a r}{\eta t^2} \exp(-r^2/4\eta t),   \label{eq_taylor}
\end{equation}
with some constant $a$ and the ``age'' of the vortex $t$.
A least-squares fit of Eq.~(\ref{eq_taylor}) to our measured curve $v_t(r)$ describes the $v_t$ profile for the average inflow surprisingly well. We can use $4\eta t \approx 7.8~$Mm from the fit to obtain a crude estimate of the turbulent viscosity. By identifying the vortex age $t$ with the supergranule lifetime (${\sim}1~$day), we get $\eta \sim (1~\text{day}/t)\times 180$~km$^2$~s$^{-1}$. This is similar to values from the literature. For example, \citet{duvall_2000} and \citet{simon_1997} obtained $\eta = 250$~km$^2$~s$^{-1}$ using helioseismology and local correlation tracking of granules, respectively. The order of magnitude of our estimate for $\eta$ also agrees with previous measurements of the diffusion coefficient of small magnetic elements \citep[and references therein]{jafarzadeh_2014}.


\section{Summary}

\subsection{Validation}
We have successfully measured the horizontal divergence and vertical vorticity of near-surface flows in the Sun using different techniques (TD and LCT), as well as different instruments (HMI and MDI). Horizontal flow velocities from LCT compare well with line-of-sight Dopplergrams (correlation coefficient 0.94).
Horizontal divergence maps from TD and LCT are in excellent agreement for 8-h averaging (correlation coefficient 0.96 for $75 \lesssim kR_\odot\lesssim 175$). Vertical vorticity measurements from TD and LCT are highly correlated at large spatial scales (correlation coefficient larger than 0.7 for $kR_\odot\le100$).

We studied the average properties of supergranules by averaging over 3\,000 of them in latitude strips from $-60\degr$ to $60\degr$. The vertical vorticity maps as measured from HMI TD and HMI LCT for the average supergranule agree at low and mid latitudes.
Above $\pm40\degr$ latitude, however, the LCT and TD results are different, due to geometrical center-to-limb systematic errors. After correcting for these errors using measurements at the equator away from the central meridian \citep[cf.][]{zhao_2013}, TD and LCT results agree well.
For MDI, the TD maps are dominated by systematic errors even at low latitudes. Therefore, HMI is a significant improvement over MDI.

\subsection{Scientific results: spatial maps of vertical vorticity}
Our findings can be summarized as follows.
The root mean square of the vertical vorticity in a map of size ${\sim}180 \times 180$~Mm$^2$ at the equator and 8~h averaging is about $15\times 10^{-6}~$s$^{-1}$ after low-pass filtering (power at scales $kR_\odot <300$).

After averaging over several thousand supergranules, the average outflow and inflow regions possess a net vertical vorticity (except at the equator). The latitudinal dependence of the vorticity magnitude is consistent with the action of the Coriolis force: $\omega_z(\lambda) \propto \Omega(\lambda) \sin \lambda / \Omega_0$. In the northern hemisphere, horizontal outflows are associated with clockwise motion, whereas inflows are associated with anti-clockwise motion. In the southern hemisphere, the sense of rotation is reversed. This resembles the behavior of high and low pressure areas in the Earth's weather system (e.g., hurricanes).

Vortices in the average supergranular inflow regions are stronger and more localized than in outflow regions. For example, at $40\degr$ latitude, the vertical vorticity is $8\times 10^{-6}~$s$^{-1}$ anti-clockwise in inflows versus $4\times 10^{-6}~$s$^{-1}$ clockwise in outflows, whereas the FWHM is 8~Mm versus 13~Mm.
The maximum tangential velocity in the average vortex is about 12~m~s$^{-1}$ at $\pm40\degr$ latitude, which is about 26 and 18 times smaller than the maximum radial flow component for outflow and inflow regions, respectively.

We have demonstrated the ability of TD and LCT to characterize rotating convection near the solar surface. This information can be used in the future to constrain models of turbulent transport mechanisms in the solar convection zone \citep[cf., e.g.,][]{ruediger_2014}. The azimuthally averaged velocity components $v_r$ and $v_t$ for supergranular outflows and inflows at various latitudes are available as online data.

\begin{acknowledgements}
JL, LG, and ACB designed research. JL performed research, analyzed data, and wrote the paper. JL and LG acknowledge research funding by Deutsche Forschungsgemeinschaft (DFG) under grant SFB 963/1 ``Astrophysical flow instabilities and turbulence'' (Project A1).
The HMI data used are courtesy of NASA/SDO and the HMI science team.
The data were processed at the German Data Center for SDO (GDC-SDO), funded by the German Aerospace Center (DLR). We thank J. Schou, T.~L. Duvall Jr., and R. Cameron for useful discussions. We are grateful to R. Burston and H. Schunker for providing help with the data processing, especially the tracking and mapping. We used the workflow management system Pegasus (funded by The National Science Foundation under OCI SI2-SSI program grant \#1148515 and the OCI SDCI program grant \#0722019).
\end{acknowledgements}

\bibliographystyle{aa}
\bibliography{literature}

\Online

\begin{appendix}

\section{Ridge filters} \label{sect_filters}
Prior to the travel-time measurements, the wavefield that is present in the Dopplergrams is filtered to select single ridges (the f modes or the p$_1$ modes). The goal is to capture as much of the ridge power as possible, even if the waves are Doppler-shifted due to flows. At the same time, we want to prevent power from neighboring ridges from leaking in and select as little background power as possible.

To construct the filter, we first measure the power spectra of the Dopplergrams at the equator and averaged over 60 days (59 days) of data in the case of HMI (MDI). After further azimuthal averaging, we identify the frequency $\omega_\text{mode}$ where the ridge maximum is located as a function of wavenumber $k$.

The filter is constructed for each $k$ as a plateau of width $2\omega_\delta$ centered around the ridge maximum $\omega_\text{mode}$. The lower and upper boundaries of the plateau we call $\omega_b$ and $\omega_c$. Next to the plateau, we add a transition region of width $\omega_\text{slope}$, which consists of a raised cosine function that guides the filter from one to zero, symmetrically around $\omega_\text{mode}$. The lower and upper limits of the filter we call $\omega_a$ and $\omega_d$, respectively.

The plateau half-width $\omega_\delta$ consists of the following terms
\begin{equation}
 \omega_\delta(k) = \frac{\omega_\Gamma(k)}{2} + \omega_v(k) + \omega_\text{const},
\end{equation}
where $\omega_\Gamma$(k) is the FWHM of the ridge (measured from the average power spectra), $\omega_v(k) = akv_\text{max}$ is the Doppler shift due to a hypothetical flow of magnitude $v_\text{max}$ multiplied by a scale factor $a$, and $\omega_\text{const}$ is a constant term of small magnitude that broadens the filter predominantly at small wavenumbers.

The width of the transition region relative to the plateau width is
\begin{equation}
 \omega_\text{slope} = j \omega_\delta ,
\end{equation}
where $j$ is a unitless factor.

In addition, we restrict the filter to a range of wavenumbers. Above and below a $k$ interval, the filters are set to zero. The $k$ limits of the interval are chosen such that the ridge power is roughly twice the background power. Because $\omega_\text{mode}$ is a function of wavenumber, these limits can also be expressed as frequencies $\omega_\text{min}$ and $\omega_\text{max}$.

Table \ref{table_filter_parameters} lists the filter parameters we chose for the f-mode and p$_1$-mode ridge filters that we use throughout the paper. Note that we use the same filters for all latitudes and longitudes. For the p$_1$ modes, we also list an alternative filter that we use to discuss the impact of the filter details on the travel-time measurements (see Appendix~\ref{sect_filter-impact}).

   \begin{table}
     \caption{Parameters of the ridge filters that are used for the travel-time measurements in this paper (see text for details). The lower part of the table gives the filter limits at $kR_\odot = 800$. The filter limits for HMI and MDI are equivalent.}             
\label{table_filter_parameters}    
\centering                     
\begin{tabular}{c c c c}    
\hline\hline               
Parameter & \multicolumn{3}{c}{Selected ridge}  \\
 & f modes & p$_1$ modes & p$_1$ modes \\
 &         & (regular)  & (alternative) \\
\hline              
 $\omega_\text{min}/2\pi$ & 1.75~mHz & 1.90~mHz & 1.90~mHz \\
 $\omega_\text{max}/2\pi$ & 5.00~mHz & 5.40~mHz & 5.00~mHz \\
 $\omega_\text{const}/2\pi$ & 0.025~mHz & 0.025~mHz & 0.030~mHz \\
 $v_\text{max}$  & 500~m~s$^{-1}$ & 500~m~s$^{-1}$ & 500~m~s$^{-1}$ \\
 $a$   & 1.0 & 1.0 & 2.0 \\
 $j$   & 1.0 & 1.0 & 0.6 \\
\hline
 $\omega_a/2\pi$ & 2.50~mHz & 3.10~mHz & 3.03~mHz \\
 $\omega_b/2\pi$ & 2.67~mHz & 3.29~mHz & 3.20~mHz \\
 $\omega_\text{mode}/2\pi$ & 2.84~mHz & 3.48~mHz & 3.48~mHz \\
 $\omega_c/2\pi$ & 3.01~mHz & 3.67~mHz & 3.77~mHz \\
 $\omega_d/2\pi$ & 3.18~mHz & 3.86~mHz & 3.94~mHz \\
\hline
\end{tabular}
   \end{table}

\section{Conversion of travel times into flow velocities} \label{chap_conversion}
Point-to-point travel times $\tau^\text{diff}(\vec{r}_1,\vec{r}_2)$ are sensitive to flows in the direction of $\vec{r}_1 - \vec{r}_2$. If the flow structure is known, travel times $\tau^\text{diff}$ can be predicted with the knowledge of sensitivity kernels. Conversely, the velocity field can be obtained from measured travel times by an inversion. Such inversion processes are, however, delicate to solve due to the illposedness of the problem. A simple way to obtain rough estimates of the flow velocity while avoiding inversions is the multiplication of the travel times by a constant conversion factor. Such a conversion factor can be calculated by artificially adding the signature of a uniform flow of known magnitude and direction to Dopplergrams. The magnitude of the measured travel time divided by the input flow speed yields the conversion factor. In the following, we describe this process.

First, we create data cubes $\phi_v(\vec{r},t)$ that have Doppler-shifted power spectra to mimic the effect of a flow $\vec{v}$ independent of position $\vec{r}$ and time $t$. The data cubes are based on the noise model by \citet{gizon_2004}, so signatures from flows others than $\vec{v}$ are not present. Following the noise model, we construct in Fourier space $\phi_v(\vec{k},\omega) = \sqrt{\mathcal{P}_v(\vec{k},\omega)} \mathcal{N}_{0,1}(\vec{k},\omega)$. Here $\vec{k}$ is the horizontal wave vector, $\mathcal{P}_v$ is a Doppler-shifted power spectrum and, at each ($\vec{k},\omega$), $\mathcal{N}_{0,1}$ are independent complex Gaussian random variables with zero mean and unit variance. Employing $\mathcal{N}_{0,1}$ ensures that the values $\phi_v(\vec{k},\omega)$ are uncorrelated, which means that there is no signal from wave scattering. We use $\mathcal{P}_v(\vec{k},\omega) = \mathcal{P}_0(\vec{k},\omega-\delta\omega)$ based on an average power spectrum $\mathcal{P}_0$ that was measured from 60 days of HMI Dopplergrams (and 59 days of MDI Dopplergrams) at the solar equator. The quantity $\delta\omega = \vec{k} \cdot \vec{v}$ is the frequency shift due to a background flow $\vec{v} = (v_x,0)$ that we add.
We construct 8~h datasets $\phi_v(\vec{r},t)$ for $v_x$ in the range between $-1\,000$ and $1\,000$~m~s$^{-1}$ in steps of $100$~m~s$^{-1}$. For each velocity value, we compute 10 realizations.

As a consistency check, we apply a second method for adding an artificial velocity signal to the HMI Dopplergram datasets. This procedure consists of tracking at an offset rate. The tracking parameters from \citet{snodgrass_1984} are modified by a constant corresponding to a $v_x$ velocity of $-100$~m~s$^{-1}$ and $100$~m~s$^{-1}$, respectively. The tracking and mapping procedure is as for the regular HMI observations. We produce 112 such datacubes for each $v_x$ value at the solar equator.

For both methods, the 8~h datasets are ridge-filtered like the normally tracked Doppler observations (f modes and p$_1$ modes). We measure travel times $\tau^\text{diff}$ in the $x$ direction with the pairs of measurement points separated by 10~Mm. This distance matches the separation in the $\tau^\text{ac}$ measurements. The reference cross-covariance $C^\text{ref}$ is taken from the regularly tracked HMI (MDI) observations averaged over 60 days (59 days) of data at the solar equator. This ensures that the artificial flow signal is captured by the travel-time measurements.

   \begin{figure*}
  \sidecaption
\includegraphics[width=0.35\hsize]{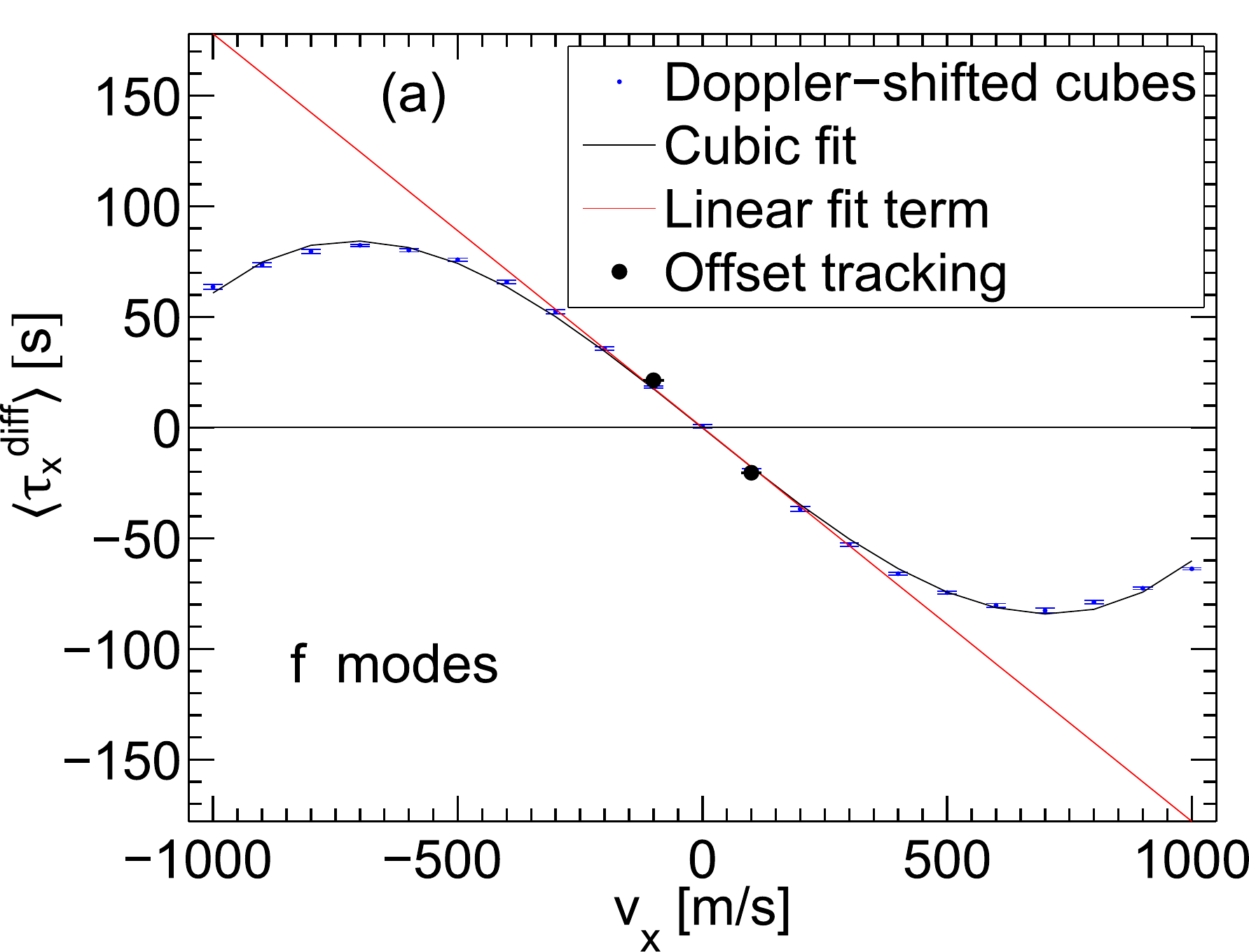}
\includegraphics[width=0.35\hsize]{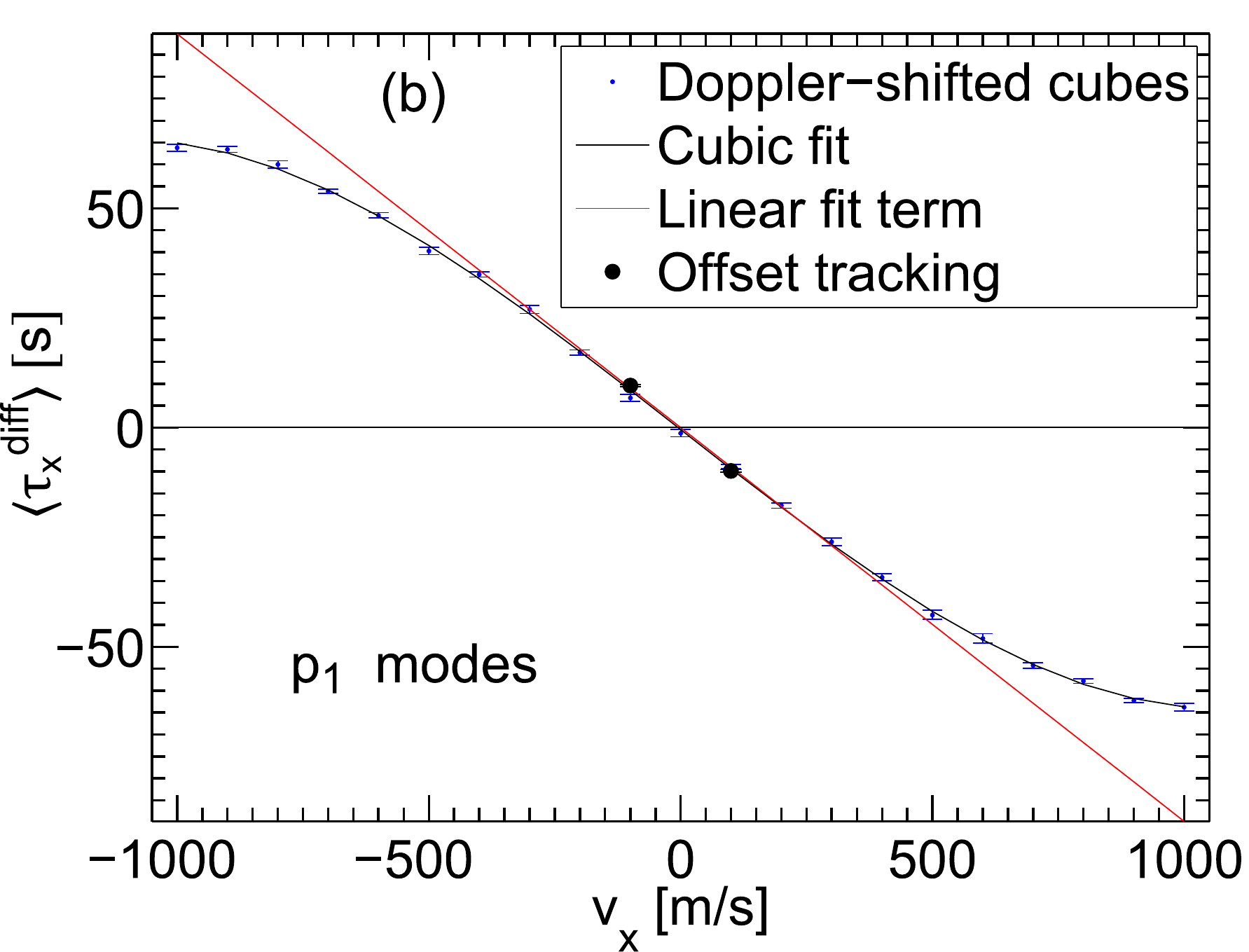}
   \caption{Point-to-point travel times from HMI Dopplergrams with artificial velocity signal. The point separation is 10~Mm in the east-west direction. \textbf{a)} f modes. \textbf{b)} p$_1$ modes. The blue dots give the travel times from Dopplergram series that were constructed using the noise model by \citet{gizon_2004}. We applied a least-squares fit with a polynomial of degree three to the resulting data (\textbf{black} curves). The red curves show the linear term of the fit. For comparison, the black filled circles show travel times from HMI Dopplergrams that were tracked at an offset rate.}
\label{fig_conversion}
    \end{figure*}

The resulting $\tau^\text{diff}$ values averaged over maps and datasets are shown for HMI in Fig.~\ref{fig_conversion}. For both f and p$_1$ modes, the travel times from offset tracking are systematically larger than for the Doppler-shifted power spectra by about 10 to 15\%. In general, the travel-time magnitudes are larger for the f modes than for the p$_1$ modes for the same input velocity value. The relation between input velocity $v_x$ and output travel time $\tau^\text{diff}$ is linear only in a limited velocity range. Whereas this range spans from roughly $-700$~m~s$^{-1}$ to $700$~m~s$^{-1}$ for the p$_1$ modes, it only reaches from $-200$~m~s$^{-1}$ to $200$~m~s$^{-1}$ for the f modes. For velocity magnitudes larger than $700$~m~s$^{-1}$, the measured f-mode travel times even decrease. However, the supergranular motions that we analyze reach typical velocities of ${\sim}300$~m~s$^{-1}$, which is well below that regime.

We applied a least-squares fit to a polynomial of degree three to the $\tau^\text{diff}$ measurements from Doppler-shifted cubes (pink curve):
\begin{equation}
 \tau^\text{diff}_x(v_x) = h_3 v_x^3 + h_2 v_x^2 + h_1 v_x + h_0 .    \label{eq_polynomial}
\end{equation}
The linear term of the polynomial is shown for HMI as the red curve in Fig.~\ref{fig_conversion}. For the actual conversion, only the linear coefficient $h_1$ is used. We obtain $h_1 = -0.178$~s$^2$~m$^{-1}$ for the f modes and $h_1 = -0.090$~s$^2$~m$^{-1}$ for the p$_1$ modes. For comparison, the coefficients $h_1$ are listed for different distances in Table~\ref{table_conversion}. The table also contains the coefficients for MDI. We convert travel times into velocities by multiplying the travel times by $1/h_1$. The velocities obtained from converting $\tau^\text{ac}$ maps we call $v^\text{ac}$.

   \begin{table}
     \caption{Coefficients $h_1$ of the cubic polynomial defined in Eq.~(\ref{eq_polynomial}) obtained from a least-squares fit. The coefficient $h_1$ for a distance of 10~Mm is used to convert measured travel times into flow velocities. For comparison, the coefficients for other distances are also given.} 
\label{table_conversion}   
\centering                      
\begin{tabular}{c c c c}     
\hline\hline           
Instrument & Distance & \multicolumn{2}{c}{Value of coefficient $h_1$ [s$^2$~m$^{-1}$] for}  \\
 & & f modes & p$_1$ modes \\
\hline              
HMI & 5~Mm & $-9.61\times 10^{-2}$ & $-4.48\times 10^{-2}$ \\
    & 10~Mm & $-1.78\times 10^{-1}$ & $-8.98\times 10^{-2}$ \\
    & 15~Mm & $-2.20\times 10^{-1}$ & $-1.39\times 10^{-1}$ \\
    & 20~Mm & $-2.15\times 10^{-1}$ & $-1.71\times 10^{-1}$ \\
\hline
MDI & 5~Mm & $-7.34\times 10^{-2}$ & $-4.66\times 10^{-2}$ \\
    & 10~Mm & $-1.58\times 10^{-1}$ & $-8.91\times 10^{-2}$ \\
    & 15~Mm & $-1.97\times 10^{-1}$ & $-1.12\times 10^{-1}$ \\
    & 20~Mm & $-2.43\times 10^{-1}$ & $-1.58\times 10^{-1}$ \\
\hline  
\end{tabular}
   \end{table}

\section{Systematic errors} 
\subsection{Center-to-limb systematics}  \label{sect_center-to-limb-systematics}
At high latitudes, the original $v^\text{ac}$ and LCT $\omega_z$ maps for the average supergranule show strong deviations from the azimuthally symmetric peak-ring structures that are visible at low latitudes. Considering that the magnitude of $\tau^\text{ac}$ and $\omega_z$ is much smaller than the magnitude of $\tau^\text{oi}$ and $\text{div}_h$ at any latitude, it is possible that even a small anisotropy in the divergent flow component of the average supergranule is picked up by the $v^\text{ac}$ and $\omega_z$ measurements and added to the signal from the tangential flow component that we want to measure. Such anisotropies can arise from various origins. Among them are geometrical effects that depend on the distance to the disk center.

For TD measurements, the sensitivity kernels depend on the distance to the limb. At $60\degr$ off disk center, $\tau^\text{diff}$ sensitivity kernels for measurements in the direction along the limb differ strongly from kernels for measurements in the center-to-limb direction \citep[see, e.g.,][for a discussion]{jackiewicz_2007}. Additionally, there is a gradient of the root mean square travel time in the center-to-limb direction.

   \begin{figure*}
  \sidecaption
\includegraphics[width=12.95cm]{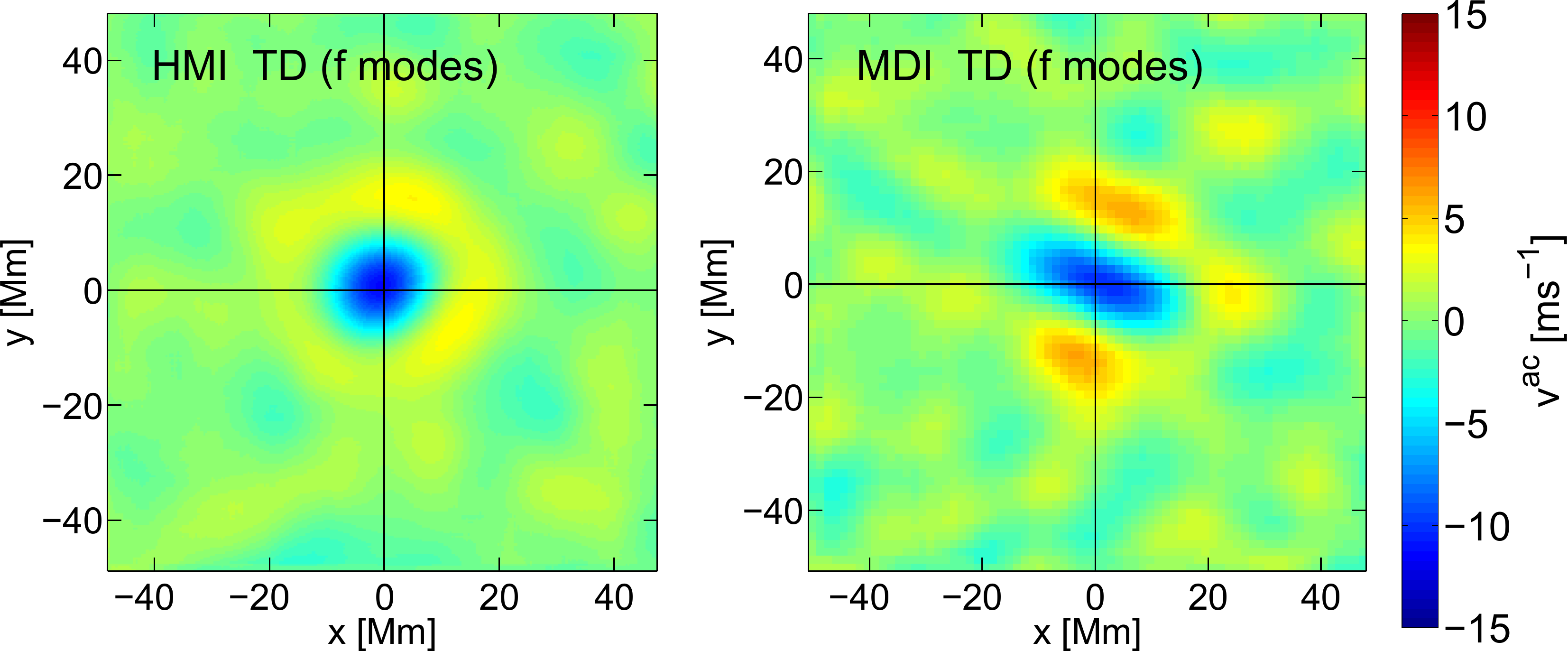}
\caption{Circulation velocities $v^\text{ac}$ of the average supergranule outflow region at solar latitude $40\degr$ derived from HMI and MDI Dopplergrams (after the correction for center-to-limb systematics). The velocity maps have been obtained by applying the respective conversion factors from Appendix~\ref{chap_conversion} to the travel times $\tau^\text{ac}$. The limits of the colorscale are arbitrarily set to $\pm15$~m~s$^{-1}$.}
\label{fig_MDI_outflow}
    \end{figure*}

   \begin{figure*}
  \sidecaption
\includegraphics[width=0.35\hsize]{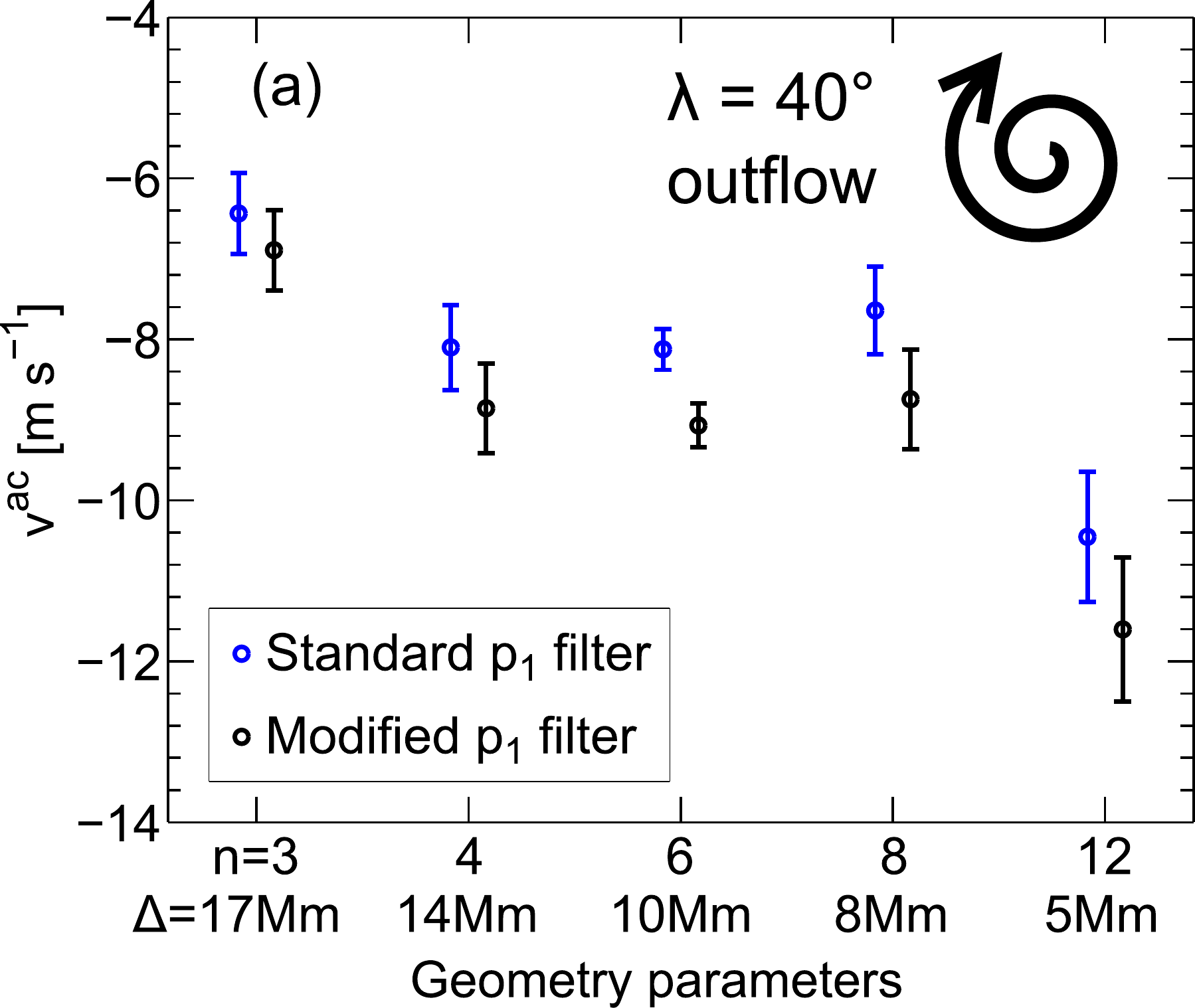}
\includegraphics[width=0.35\hsize]{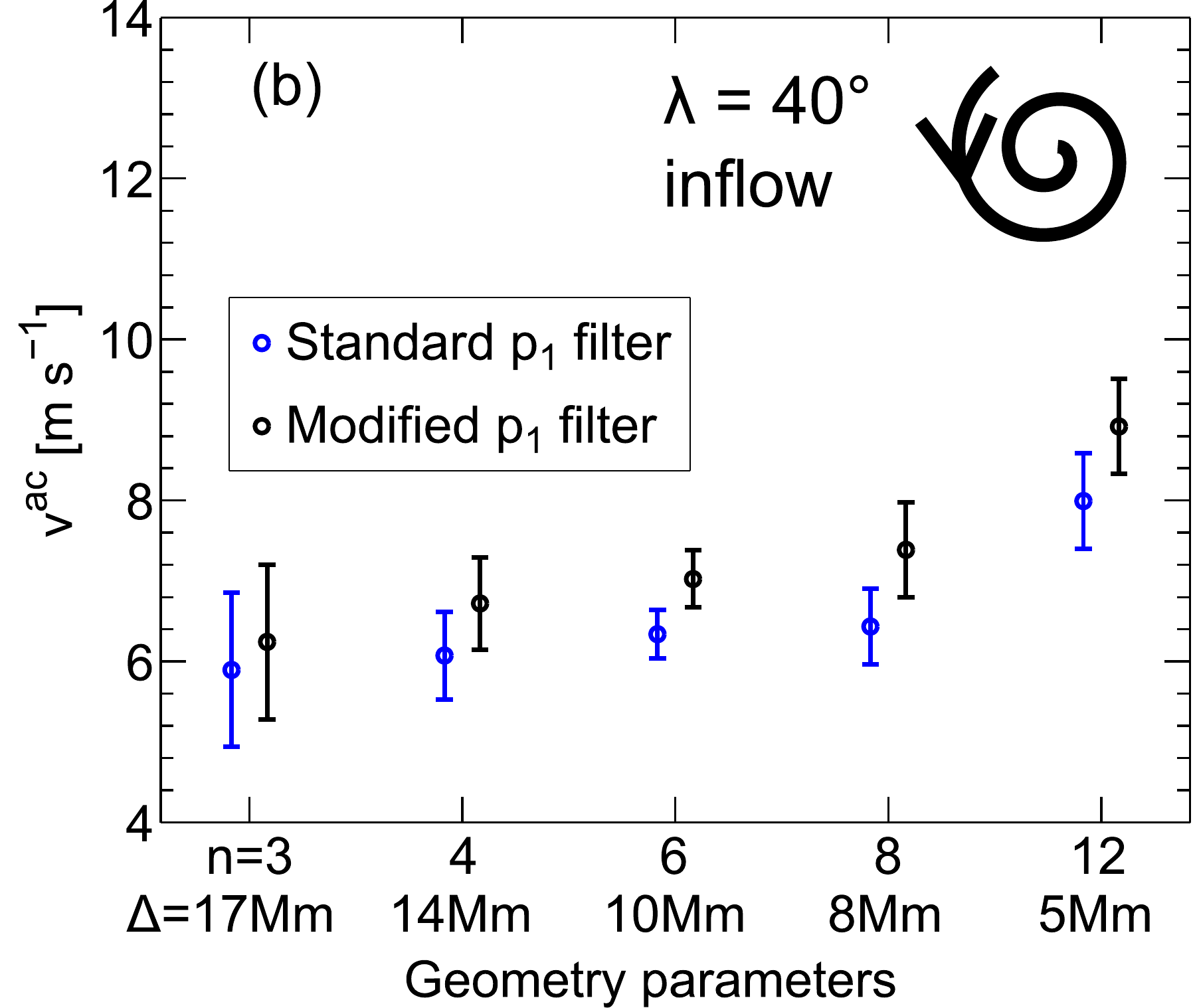}
   \caption{Peak $v^\text{ac}$ values for p$_1$ modes using different parameter combinations ($\Delta,n$) for the average supergranule at solar latitude $40\degr$. \textbf{a)} In the average outflow region. \textbf{b)} In the average inflow region. The blue symbols give the results for the p$_1$ ridge filter that has been used throughout this paper. For the results in black, an alternative p$_1$ ridge filter with slightly different parameters was used (see text for details). The errorbars have been computed as in Fig.~\ref{fig_aligned_vs_lat}. The annulus radii corresponding to the various combinations ($\Delta,n$) are all within ($10.0\pm0.5$)~Mm.}
\label{fig_p1comparison}
    \end{figure*}

In the case of LCT, the shrinking Sun effect causes large-scale gradients of the horizontal velocity (of several hundred meters per second) pointing towards disk center \citep{lisle_2004}. This effect is presumably caused by insufficient resolution of the granules. Although HMI intensity and Doppler images have a pixel size of about 350~km at disk center, the point spread function has a FWHM of about twice that value. In Dopplergrams, the hot, bright, and broad upflows in the granule cores cause stronger blueshifts than the redshifts from the cooler, darker, and narrow downflows. Due to the insufficient resolution, the granules appear blue-shifted as a whole. This blue shift adds to the blue shift of granules that move toward the observer (i.e., towards disk center), giving them a stronger signal in the Dopplergram. \citeauthor{lisle_2004} argue that LCT of Dopplergrams gives more weight to these granules than to those granules that move away from the observer. However, it is not clear what causes the shrinking Sun effect in LCT of intensity images. Fortunately, the shrinking Sun effect appears to be a predominantly large-scale and time-independent effect, so it can easily be removed from LCT velocity maps by subtracting a mean image.

Another problem is the foreshortening. Far away from the disk center, the granules are not resolved as well in the center-to-limb direction as in the perpendicular horizontal direction. This introduces a dependence of the measurement sensitivity on angle. Indeed we measure at $\pm60\degr$ latitude that the radial flow component $v_r$ of the average supergranule is weaker by 15 to 20\% in the center-to-limb direction compared to the perpendicular direction. This corresponds to a maximum velocity difference of about 50~m~s$^{-1}$ for outflows and 30~m~s$^{-1}$ for inflows.
At $40\degr$ latitude, in contrast, this difference is less than 2\% (6~m~s$^{-1}$).

\subsection{MDI instrumental systematics}
Whereas for HMI the removal of geometrical center-to-limb effects results in similar $v^\text{ac}$ peak structures in the supergranule outflow regions in the whole latitude range from $-60\degr$ to $60\degr$, for MDI the peak structures appear asymmetric and distorted even after the correction. An example for f-mode TD at $40\degr$ latitude is shown in Fig.~\ref{fig_MDI_outflow}.
Even at disk center where geometrical effects should not play a role, there are visible systematic features (that do not appear for HMI, cf. Fig.~\ref{fig_aligned-ccw-outflow}). This is probably due to instrumental effects that are specific to MDI \citep[see, e.g.,][for a discussion of instrumental errors in MDI]{korzennik_2004}.

\subsection{Selection of filter and $\tau^\text{ac}$ geometry parameters} \label{sect_filter-impact}
We note that the $v^\text{ac}$ velocity results for TD depend on the details of the ridge filter as well as the geometry parameters $(\Delta,n)$ of the $\tau^\text{ac}$ measurements.

To give an idea of this, we construct an alternative p$_1$ ridge filter with slightly different width parameters (see Appendix \ref{sect_filters}).
Additionally, we select four other combinations $(\Delta,n)$ of $\tau^\text{ac}$ measurements that preserve the annulus radius $R$, so that $R$ is within ($10.0\pm0.5$)~Mm for all the combinations $(\Delta,n)$. As for the standard combination $(\Delta=10~$Mm$,n=6)$, we use four different angles $\beta$ for each additional combination.

For all these combinations and both the standard and modified p$_1$ filters, we calculated $v^\text{ac}$ for the average supergranule at 40$\degr$ latitude. The resulting peak velocities are shown in Fig.~\ref{fig_p1comparison} for both inflow and outflow regions. Note that we did not apply the center-to-limb correction since it only has a weak influence on the peak velocity magnitude at $40\degr$ latitude.

Evidently, the modified p$_1$ filter results in systematically larger $v^\text{ac}$ amplitudes. The difference with respect to the standard filter increases with decreasing $\Delta$. For $\Delta = 10~$Mm and $n=6$, it is about 10\%.
This is qualitatively in line with \citet{duvall_2013}. Using phase-speed filters, \citeauthor{duvall_2013} observed that the strength of the travel-time signal from supergranulation is strongly dependent on the filter width. This shows that one should be careful when comparing absolute velocities from TD and LCT. For more reliable velocity values, an inversion of $\tau^\text{oi}$ and $\tau^\text{ac}$ maps would be needed.

The comparison of different combinations $(\Delta,n)$ for the same filter shows that for $n=4$, 6, and 8 the $v^\text{ac}$ amplitudes are similar, so selecting the combination $(\Delta=10~$Mm$,n=6)$, as we did for most of this work, appears justified. Decreasing $\Delta$ to about 5~Mm changes the peak $v^\text{ac}$ values. A possible reason is that $\Delta$ in this case becomes comparable to the wavelength of the oscillations, so it is harder to distinguish between flows in opposite directions. For small $n$, on the other hand, the measurement geometry deviates strongly from a circular contour. This might explain the deviations in $v^\text{ac}$ for $n=3$.

\end{appendix}

\end{document}